\documentclass[12pt,onecolumn]{IEEEtran}

\ifCLASSINFOpdf
   \usepackage[pdftex]{graphicx}

\else

\fi

\usepackage[cmex10]{amsmath}
\usepackage{amssymb}   
\usepackage{amsxtra}
\usepackage{amscd}
\usepackage{amsthm}

\usepackage{graphicx}   

\usepackage{array}  

\usepackage{multirow}  

\usepackage{cite}

\usepackage{color,soul} 
\soulregister\cite7
\soulregister\ref7
\soulregister\pageref7

\begin{document}
%
\title{{\huge On Multiple-Input Multiple-Output OFDM with Index Modulation for Next Generation Wireless Networks}}


%
%
%

\author{Ertugrul~Basar,~\IEEEmembership{Senior Member,~IEEE} 
	\thanks {Copyright (c) 2016 IEEE. Personal use of this material is permitted. However, permission to use this material for any other purposes must be obtained from the IEEE by sending a request to pubs-permissions@ieee.org.}
	\thanks{The author is with Istanbul Technical University, Faculty of Electrical and Electronics Engineering, 34469, Istanbul, Turkey. e-mail: basarer@itu.edu.tr.} 
    \thanks{Digital Object Identifier 10.1109/TSP.2016.2551687}
	}

%
%

\markboth{IEEE Transactions on Signal Processing}{Ba\c{s}ar:  On Multiple-Input Multiple-Output OFDM with Index Modulation for Next Generation Wireless Networks}

%




\maketitle

\begin{abstract}
Multiple-input multiple-output orthogonal frequency division multiplexing with index modulation (MIMO-OFDM-IM) is a novel multicarrier transmission technique which has been proposed recently as an alternative to classical MIMO-OFDM. In this scheme, OFDM with index modulation (OFDM-IM) concept is combined with MIMO transmission to take advantage of the benefits of these two techniques. In this paper, we shed light on the implementation and error performance analysis of the MIMO-OFDM-IM scheme for next generation 5G wireless networks. Maximum likelihood (ML), near-ML, simple minimum mean square error (MMSE) and ordered successive interference cancellation (OSIC) based MMSE detectors of MIMO-OFDM-IM are proposed and their theoretical performance is investigated. It has been shown via extensive computer simulations that MIMO-OFDM-IM scheme provides an interesting trade-off between error performance and spectral efficiency as well as it achieves considerably better error performance than classical MIMO-OFDM using different type detectors and under realistic conditions.

\end{abstract}

%
\IEEEpeerreviewmaketitle

\renewcommand{\thefootnote}{\arabic{footnote}}

\section{Introduction}
%
%
%
%
\IEEEPARstart{O}{rthogonal} frequency division multiplexing (OFDM)  is one of the most popular multi-carrier transmission techniques to satisfy the increasing demand for high data rate wireless communications systems. OFDM technologies have become an integral part of many standards such as Long Term Evolution (LTE), IEEE 802.11x wireless local area network (LAN), digital video broadcasting (DVB) and IEEE 802.16e-WiMAX due to their efficient implementation and robustness to inter-symbol interference.

Multiple-input multiple-output (MIMO) transmission techniques have been widely studied over the past decade due to their advantages over single antenna systems such as improved data rate and energy efficiency. Spatial modulation (SM), which is based on the transmission of information bits by means of the indices of the active transmit antennas of a MIMO system \cite{SM_jour}, is one of the promising MIMO solutions towards spectral and energy-efficient next generation communications systems \cite{5G2}. SM has attracted significant attention by the researchers over the past few years \cite{SSK,GSM2,STBC_SM,SM_imperfect,SM_BEP,SM_practical} and it is still a hot topic in wireless communications \cite{SM_magazine_2}.

OFDM with index modulation (OFDM-IM) \cite{OFDM_IM} is a novel multicarrier transmission technique which has been proposed as an alternative to classical OFDM. Inspiring from the SM concept, in OFDM-IM, index modulation techniques are applied for the indices of the available subcarriers of an OFDM system. In OFDM-IM scheme, only a subset of available subcarriers are selected as active according to the information bits, while the remaining inactive subcarriers are set zero. In other words, the information is transmitted not only by the data symbols selected from $M$-ary signal constellations, but also by the indices of the active subcarriers. Unlike classical OFDM, the number of active subcarriers can be adjusted in the OFDM-IM scheme, and this flexibility in the system design provides an interesting trade-off between error performance and spectral efficiency. Furthermore, it has been shown that OFDM-IM has the potential to achieve a better error performance than classical OFDM for low-to-mid spectral efficiency values. Due to its adjustable number of active subcarriers, OFDM-IM can be a possible candidate not only for high-speed wireless communications systems but also for machine-to-machine (M2M) communications systems which require low power consumption.

Subcarrier index modulation concept for OFDM \cite{OFDM_IM,SIM_OFDM,ESIM_OFDM} has attracted significant attention from the researchers over the past two years and it has been investigated in some very recent studies \cite{MCIK_OFDM,OFDM_GIM,Opt_OFDM_IM,GSScM,OFDM_ISIM,IM_OFDM_V2X,CI_OFDM_IM,Wen1,Wen2,MIMO_OFDM_IM}. A tight approximation for the error performance of OFDM-IM is given in \cite{MCIK_OFDM}. By the selection of active subcarriers in a more flexible way to further increase the spectral efficiency, OFDM-IM scheme is generalized in \cite{OFDM_GIM}. The problem of the selection of optimal number of active subcarriers is investigated in \cite{Opt_OFDM_IM} and \cite{GSScM}. In \cite{OFDM_ISIM}, subcarrier level block interleaving is introduced for OFDM-IM in order to improve its error performance by taking advantage of uncorrelated subcarriers. In \cite{IM_OFDM_V2X}, OFDM-IM with interleaved grouping is adapted to vehicular communications. OFDM-IM is combined with coordinate interleaving principle in \cite{CI_OFDM_IM} to obtain additional diversity gains. More recently, it has been proved that OFDM-IM and its variants outperform the classical OFDM in terms of ergodic achievable rate\cite{Wen1} and coding gain\cite{Wen2}.

Considering the advantages of OFDM and MIMO transmission techniques, the combination of them unsurprisingly appears as a strong alternative for 5G and beyond wireless networks \cite{5G}. MIMO-OFDM-IM, which is obtained by the combination of MIMO and OFDM-IM transmission techniques, is a recently proposed high-performance multicarrier transmission technology and can be considered as a possible alternative to classical MIMO-OFDM \cite{MIMO_OFDM_IM}. In this scheme, each transmit antenna transmits its own OFDM-IM frame to boost the data rate and at the receiver side, these frames are separated and demodulated using a novel sequential minimum mean square error (MMSE) detector which considers the statistics of the MMSE filtered received signals. However, since different applications have different error performance and decoding complexity constraints, the design and analysis of different type of detectors remain an open problem for the MIMO-OFDM-IM scheme.  

In this paper, we deal with the implementation and error performance analysis of the MIMO-OFDM-IM scheme for different type of detectors and active indices selection methods under realistic conditions. First, the maximum likelihood (ML) detector of the MIMO-OFDM-IM scheme is investigated to benefit from the diversity gain of MIMO systems and its average bit error probability (ABEP) is derived by the calculation of pairwise error probability (PEP) of the MIMO-OFDM-IM subblocks. Second, in order to reduce the decoding complexity of the brute-force ML detector of the MIMO-OFDM-IM scheme, a novel low complexity near-ML detector is proposed which is shown to provide better bit error rate (BER) performance than V-BLAST type classical MIMO-OFDM for different configurations. Third, a simple MMSE detection algorithm is proposed and its theoretical ABEP is derived to shed light on the performance of MIMO-OFDM-IM for MMSE detection. Then, a novel ordered successive interference cancellation (OSIC) based sequential MMSE detector is proposed for MIMO-OFDM-IM. Finally, the error performance of  MIMO-OFDM-IM is evaluated for a realistic LTE channel model and under channel estimation errors. It has been shown via computer simulations that MIMO-OFDM-IM can be a strong alternative to classical MIMO-OFDM due to its improved BER performance and flexible system design.

The rest of the paper is organized as follows. In Section II, the system model of MIMO-OFDM-IM is presented. In Sections III and IV, we deal with ML and MMSE detection of the MIMO-OFDM-IM scheme and provide our theoretical results, respectively. Simulation results are provided in Section V. Finally, Section VI concludes the paper. 

 \textit{Notation}: Bold, lowercase and capital letters are used for column vectors and matrices, respectively. $(\mathbf{A})_{t*}$, $(\mathbf{A})_{*t}$ and $(\mathbf{A})_{t,t}$ denote the $t$th row, $t$th column and the $t$th main diagonal element of $\mathbf{A}$, respectively. $\left( \cdot \right)^*$, ${{\left( \cdot \right)}^{\mathrm{T}}}$ and ${{\left( \cdot \right)}^{\mathrm{H}}}$ denote complex conjugation, transposition and Hermitian transposition, respectively. $\det(\mathbf{A})$ and $\textrm{rank}(\mathbf{A})$ denote the determinant and rank of $\mathbf{A}$, respectively. $\mathbf{I}_{N}$ is the identity matrix with dimensions $N\times N$ and $\mbox{diag}\left(\cdot \right) $ denotes a diagonal matrix. $\left\| \cdot\right\| $ stands for the Euclidean  norm. The probability of an event is denoted by ${{P}}\left( \cdot \right)$. $E\left\lbrace \cdot \right\rbrace $ and $Var(\cdot)$ stand for expectation and variance, respectively. The covariance matrix of $\mathbf{a}$ is denoted by $\textrm{cov}(\mathbf{a})$. $X\sim \mathcal{N} \left(m_X,\sigma_X^2 \right) $ denotes the Gaussian distribution of a real random variable (r.v.) $X$ with mean $m_X$ and variance $\sigma_X^2$, while $X\sim \mathcal{CN}\left( 0,\sigma _{X}^{2} \right)$ denotes the distribution of a circularly symmetric complex Gaussian r.v. $X$ with variance $\sigma _{X}^{2}$. $Q(\cdot)$ is the tail probability of the standard Gaussian distribution. $C\left(N,K \right) $ stands for the binomial coefficient and $\lfloor\cdot\rfloor$ is the floor function. $\mathcal{S}$ denotes $M$-ary signal constellation. $ \mathbb{C} $ and $ \mathbb{R} $ denote the ring of complex and real numbers, respectively. $\Re \left\lbrace X \right\rbrace $ and $\Im\left\lbrace X \right\rbrace$ denote the real and imaginary parts of a complex variable $X$, respectively.

\section{MIMO-OFDM-IM at a Glance}

The block diagram of the MIMO-OFDM-IM transceiver \cite{MIMO_OFDM_IM} is given in Fig. 1, where the concept of OFDM-IM, which is shown in Fig. 2, is extended to a MIMO configuration. A MIMO system with $T$ transmit and $R$ receive antennas is assumed. As seen from Fig. 1, a total of $mT$ information bits enter the  MIMO-OFDM-IM transmitter for the transmission of each MIMO-OFDM-IM frame. These $mT$ bits are first split into $T$ groups and the corresponding $m$ bits are processed in each branch of the transmitter by the OFDM index modulators as shown in Fig. 2. Unlike the classical OFDM, these $m$ bits are used not only in $M$-ary modulation but also in the selection of the indices of active subcarriers and the $N_F \times 1$ OFDM-IM block
\begin{equation}
\mathbf{x}_t=\begin{bmatrix}
x_t(1) & x_t(2) &\cdots &x_t(N_F)
\end{bmatrix}^\mathrm{T},\quad t=1,2,\ldots,T
\end{equation}
is obtained in each branch of the transmitter, where $N_F$ is the size of the fast Fourier transform (FFT) and $x_t(n_f) \in \left\lbrace 0,\mathcal{S} \right\rbrace,n_f=1,2,\ldots,N_F $.

\begin{figure*}[t]
	\begin{center}
		{\includegraphics[scale=0.8]{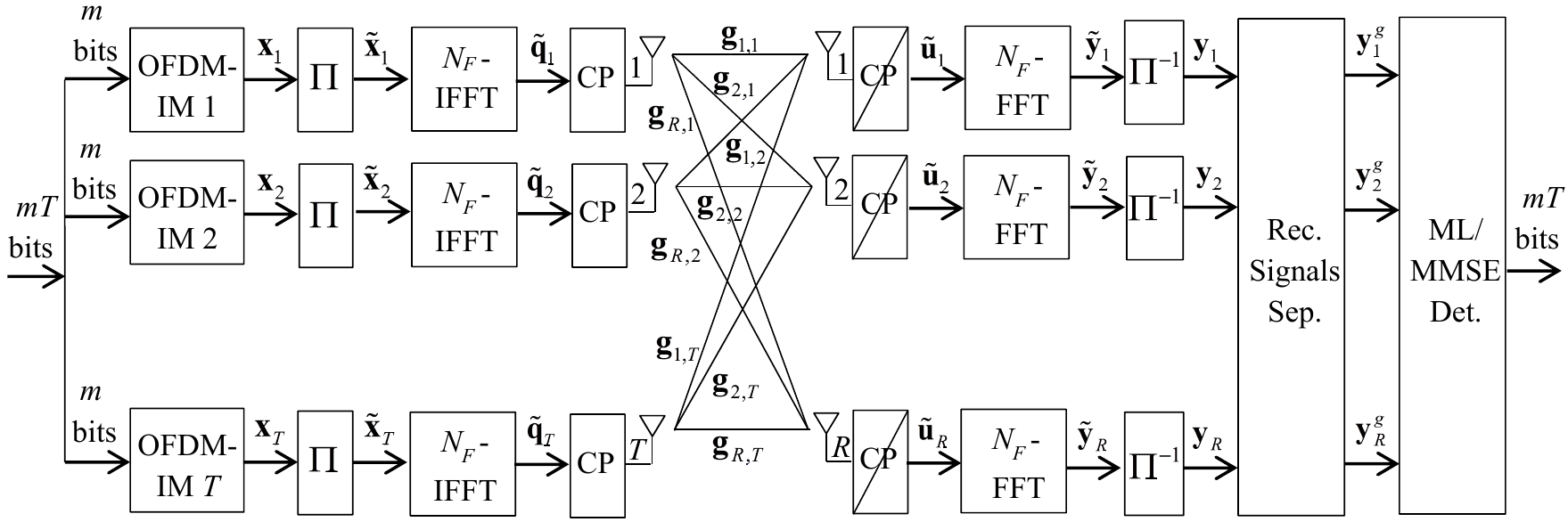}}
		\vspace*{-0.2cm}
		\caption{Transceiver Structure of the MIMO-OFDM-IM Scheme for a $T\times R$ MIMO System}
		\vspace*{-0.4cm}
	\end{center}
\end{figure*}

\begin{figure}[t]
	\begin{center}
		{\includegraphics[scale=0.8]{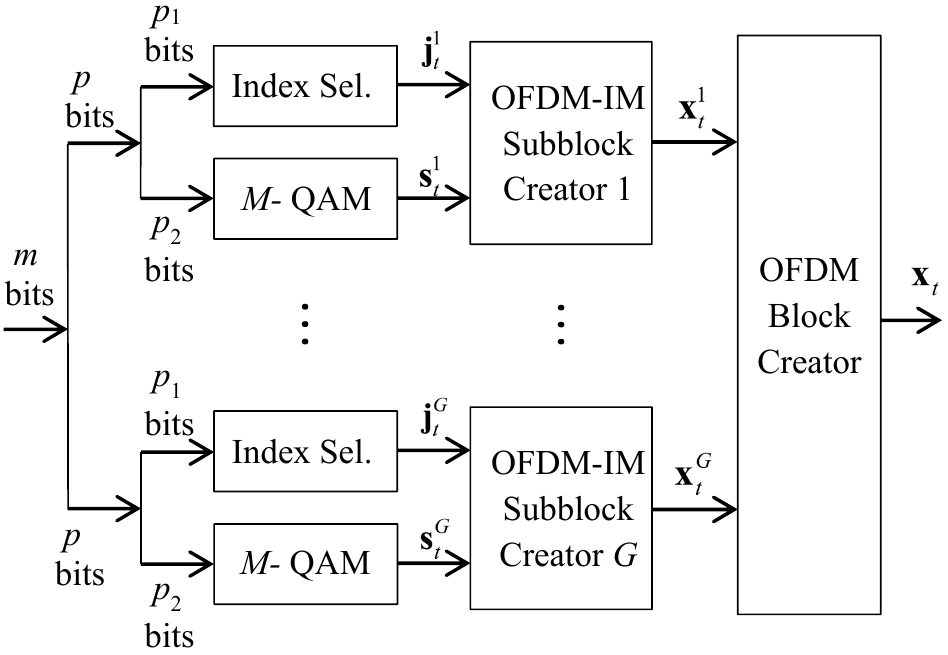}}
		\vspace*{-0.2cm}
		\caption{OFDM index modulators at each brach of the transmitter}
		\vspace*{-0.4cm}
	\end{center}
\end{figure}

According to the OFDM-IM principle \cite{OFDM_IM}, which is carried out simultaneously in each branch of the transmitter, 
these $m$ bits are split into $G$ groups each containing $p=p_1 + p_2$ bits, which are used to form OFDM-IM subblocks 
\begin{equation}
\mathbf{x}_t^g=\begin{bmatrix}
x_t^g(1) & x_t^g(2) &\cdots &x_t^g(N)
\end{bmatrix}^\mathrm{T}, \quad g=1,2,\ldots,G
\end{equation}
of length $N=N_F/G$, where $x_t^g(n) \in \left\lbrace 0,\mathcal{S} \right\rbrace,n=1,2,\ldots,N $. At each subblock $g$, considering the corresponding $p_1=\lfloor \log_2\left(C\left(N,K\right)  \right)  \rfloor$ bits, only $K$ out of $N$ available subcarriers are selected as active by the index selector, where the indices of the active subcarriers are denoted by 
\begin{equation}
\mathbf{j}^g_t=\begin{bmatrix}
j_t^g(1) & j_t^g(2) &\cdots &j_t^g(K)
\end{bmatrix}^\mathrm{T}
\end{equation}
where $j^g_t(k) \in \left\lbrace 1,\cdots,N\right\rbrace , k=1,2,\ldots,K $. On the other hand, the remaining $N-K$ subcarriers are inactive and set to zero. In the mean time, the remaining $p_2=K \log_2(M)$ bits are mapped onto the considered $M$-ary quadrature amplitude modulation ($M$-QAM) signal constellation to obtain
\begin{equation}
\mathbf{s}_t^g=\begin{bmatrix}
s_t^g(1) & s_t^g(2) &\cdots &s_t^g(K)
\end{bmatrix}^{\mathrm{T}}
\end{equation}
where $ s^g_t(k) \in \mathcal{S}, k=1,2,\ldots,K $. For each OFDM-IM subblock $\mathbf{x}_t^g$, the $K$ elements of $\mathbf{s}_t^g$ modulates the $K$ active subcarriers whose indices given by $\mathbf{j}_t^g$. Therefore, unlike classical MIMO-OFDM, $ \mathbf{x}_t,t=1,2,\ldots,T $ contains some zero terms whose positions carry information for MIMO-OFDM-IM\footnote{For the implementation of practical OFDM schemes, the number of processed subcarriers $(N_F)$ is generally chosen to be greater than the number of used subcarriers $(N_U)$, where the first and last $(N_F-N_U)/2$ elements of OFDM frames are padded with zeros. For this purpose, the first and last $(N_F-N_U)/2N$ subblocks of the MIMO-OFDM-IM scheme can be padded with zeros. However, for ease of presentation, we do not consider zero padding in this study.}.

Active subcarrier index selection is performed at OFDM index modulators of the transmitter by either using reference look-up tables for smaller values of active subcarriers $(K)$ and subblock sizes $(N)$ or an index selection procedure based on the combinatorial number theory for higher values of $K$ and  $N$. The considered reference look-up tables for $N=4,K=2$ and $N=4,K=3$ are given in Tables I and II, respectively, where $s_k\in \mathcal{S}$ for $ k=1,2,\ldots K $. As seen from Table I, for $N=4$ and $K=2$, $p_1=2$ bits can be used to determine the indices of the two active subcarriers out of four available subcarriers according to the reference look-up table of size $C=2^{p_1}=4$. For higher $K$ and $N$ values, the combinatorial algorithm provides the selected indices according to the incoming $p_1$ bits \cite{OFDM_IM}.

\begin{table}[!t]
	\begin{center}
		\setlength{\extrarowheight}{1pt}
		\caption{Reference Look-up Table for $N=4,K=2$ and $p_1=2$}
		\label{tab:Look_up}
		\begin{tabular}[c]{|c||c||c|} \hline
			\textit{Bits} & \textit{Indices $(\mathbf{j}_t^g)^\textrm{T}$} & \textit{OFDM-IM subblocks} $(\mathbf{x}_t^g)^\textrm{T}$  \\ \hline \hline
			$[0\,\,\, 0]$ & $\begin{bmatrix}1 & 3   \end{bmatrix}$ & $\begin{bmatrix}s_t^g(1) & 0 & s_t^g(2) & 0   \end{bmatrix}$  \\ \hline
			$[0\,\,\, 1]$ & $\begin{bmatrix}2 & 4   \end{bmatrix}$ & $\begin{bmatrix}0 & s_t^g(1) & 0 & s_t^g(2)   \end{bmatrix}$ \\ \hline
			$[1\,\,\, 0]$ & $\begin{bmatrix}1 & 4  \end{bmatrix}$ & $\begin{bmatrix} s_t^g(1) & 0 & 0 & s_t^g(2)     \end{bmatrix}$ \\ \hline
			$[1\,\,\, 1]$ & $\begin{bmatrix}2 & 3   \end{bmatrix}$ & $\begin{bmatrix} 0 & s_t^g(1) & s_t^g(2) &  0  \end{bmatrix}$  \\ \hline
		\end{tabular}
	\end{center}
	\vspace*{-0.3cm}
\end{table}

\begin{table}[!t]
	\begin{center}
		\setlength{\extrarowheight}{1pt}
		\caption{Reference Look-up Table for $N=4,K=3$ and $p_1=2$}
		\label{tab:Look_up2}
		\begin{tabular}[c]{|c||c||c|} \hline
			\textit{Bits} & \textit{Indices $(\mathbf{j}_t^g)^\textrm{T}$} & \textit{OFDM-IM subblocks} $(\mathbf{x}_t^g)^\textrm{T}$  \\ \hline \hline
			$[0\,\,\, 0]$ & $\begin{bmatrix}1 & 2 & 3   \end{bmatrix}$ & $\begin{bmatrix}s_t^g(1) & s_t^g(2) & s_t^g(3) & 0   \end{bmatrix}$  \\ \hline
			$[0\,\,\, 1]$ & $\begin{bmatrix}1 & 2 & 4   \end{bmatrix}$ & $\begin{bmatrix}s_t^g(1) & s_t^g(2) & 0 & s_t^g(3)   \end{bmatrix}$ \\ \hline
			$[1\,\,\, 0]$ & $\begin{bmatrix}1 & 3 & 4   \end{bmatrix}$ & $\begin{bmatrix}s_t^g(1) & 0 & s_t^g(2) & s_t^g(3)   \end{bmatrix}$ \\ \hline
			$[1\,\,\, 1]$ & $\begin{bmatrix}2 & 3 & 4   \end{bmatrix}$ & $\begin{bmatrix} 0 & s_t^g(1) & s_t^g(2) & s_t^g(3)    \end{bmatrix}$  \\ \hline
		\end{tabular}
	\end{center}
	\vspace*{-0.3cm}
\end{table}

In each branch of the transmitter, the OFDM index modulators construct the OFDM-IM subblocks first, then concatenate these $G$ subblocks to obtain the main OFDM-IM blocks $\mathbf{x}_t,t=1,2,\ldots,T$. $G\times N$ block interleavers $(\Pi)$ are employed at the transmitter to transmit the elements of the subblocks from uncorrelated channels. Then, inverse FFT (IFFT) operators process the interleaved OFDM-IM frames $\mathbf{\tilde{x}}_t,t=1,2,\ldots,T$ and obtain $\mathbf{\tilde{q}}_t,t=1,2,\ldots,T$. It is assumed that the time-domain OFDM symbols are normalized to have unit energy, i.e., $ E\left\lbrace \mathbf{\tilde{q} }_t^\textrm{H}\mathbf{\tilde{q} }_t \right\rbrace=N_F  $ for all $ t $.


After the IFFT operation, a cyclic prefix (CP) of $C_p$ samples is appended to the beginning of the OFDM-IM frames in each branch of the transmitter. After parallel-to-serial and digital-to-analog conversions, the resulting signals are sent simultaneously from $ T $ transmit antennas over a frequency-selective Rayleigh fading MIMO channel which can be represented by $\mathbf{g}_{r,t}\in \mathbb{C}^{L\times 1}$, where $L$ is the number of channel taps. We assume that the elements of $\mathbf{g}_{r,t}$ are independent and identically distributed (i.i.d.) with $\mathcal{CN}(0,1/L)$.

Based on the assumption that the wireless channels remain constant during the transmission of a MIMO-OFDM-IM frame and $C_p > L$, after removing the CP and performing FFT operations in each branch of the receiver, the input-output relationship of the MIMO-OFDM-IM scheme in the frequency domain is obtained for $r=1,2,\ldots,R$ as follows:
\begin{equation}
\mathbf{\tilde{y}}_r=\sum\nolimits_{t=1}^{T}\mbox{diag} \left( \mathbf{\tilde{x}}_t\right)  \mathbf{h }_{r,t} + \mathbf{w}_r .
\label{eq:ilk}
\end{equation}
In (\ref{eq:ilk}), $\mathbf{\tilde{y}}_r= \begin{bmatrix}
\tilde{y}_r(1) & \tilde{y}_r(2) & \cdots & \tilde{y}_r(N_F)
\end{bmatrix}^\textrm{T}$ is the vector of the received signals for receive antenna $r$, the frequency response of the wireless channel between the transmit antenna $t$ and receive antenna $r$ is denoted by $\mathbf{h}_{r,t} \in \mathbb{C}^{N_F \times 1}$, and $\mathbf{w}_{r} \in \mathbb{C}^{N_F \times 1}$ stands for the vector of noise samples. The elements of $ \mathbf{h}_{r,t} $ and $\mathbf{w}_{r}$ follow $\mathcal{CN}\left(0,1 \right) $ and $\mathcal{CN}\left(0,N_{0,F} \right) $ distributions, respectively, where $N_{0,F}$ denotes the variance of the noise samples in the frequency domain, and it is related to the variance of the noise samples in the time domain as $N_{0,F}=(K/N) N_{0,T}$.


After deinterleaving operation, the received signals for receive antenna $r$ are obtained as
\begin{equation}
\mathbf{y}_r=\sum\nolimits_{t=1}^{T}\mbox{diag} \left( \mathbf{x}_t\right)  \mathbf{\breve{h }}_{r,t} + \mathbf{\breve{w}}_r
\label{eq:1}
\end{equation}
for $ r=1,2,\ldots,R $, where $\mathbf{\breve{h }}_{r,t}$ and $\mathbf{\breve{w}}_r  $ are deinterleaved versions of $ \mathbf{h }_{r,t} $ and $\mathbf{w}_r  $, respectively. As seen from Fig. 1, before the detection of the MIMO-OFDM-IM scheme, the received signals in (\ref{eq:1}) are separated for each subblock $g=1,2,\ldots,G$ as
\begin{align}
\mathbf{y}_r &= \begin{bmatrix}
(\mathbf{y}_r^1)^{\textrm{T}} & (\mathbf{y}_r^2)^{\textrm{T}} & \cdots & (\mathbf{y}_r^G)^{\textrm{T}} 
\end{bmatrix}^{\textrm{T}} \nonumber\\
\mathbf{x}_t &= \begin{bmatrix}
(\mathbf{x}_t^1)^{\textrm{T}} & (\mathbf{x}_t^2)^{\textrm{T}} & \cdots & (\mathbf{x}_t^G)^{\textrm{T}} 
\end{bmatrix}^{\textrm{T}} \nonumber\\
\mathbf{\breve{h}}_{r,t} &= \begin{bmatrix}
(\mathbf{\breve{h}}_{r,t}^1)^{\textrm{T}} & (\mathbf{\breve{h}}_{r,t}^2)^{\textrm{T}} & \cdots & (\mathbf{\breve{h}}_{r,t}^G)^{\textrm{T}} \end{bmatrix}^{\textrm{T}} \nonumber\\
\mathbf{\breve{w}}_{r} &= \begin{bmatrix}
(\mathbf{\breve{w}}_{r}^1)^{\textrm{T}} & (\mathbf{\breve{w}}_{r}^2)^{\textrm{T}} & \cdots & (\mathbf{\breve{w}}_{r}^G)^{\textrm{T}}
\end{bmatrix}^{\textrm{T}}.
\label{3}
\end{align}
From (\ref{3}), we obtain the following signal model for each subblock $g$:
\begin{align}
\label{eq:3}
\mathbf{y}_r^g & =\sum\nolimits_{t=1}^{T}\mbox{diag} \left( \mathbf{x}_t^g\right)  \mathbf{\breve{h }}_{r,t}^g + \mathbf{\breve{w}}_r^g, \quad r=1,2,\ldots,R, 
\end{align}
where 
\begin{equation}
\mathbf{y}_r^g= \begin{bmatrix}
y_r^g(1) & y_r^g(2) & \cdots & y_r^g(N)
\end{bmatrix}^\textrm{T}
\end{equation}
is the vector of the received signals at receive antenna $r$, 
\begin{equation}
\mathbf{x}_t^g=\begin{bmatrix}
x_t^g(1) & x_t^g(2) &\cdots &x_t^g(N)
\end{bmatrix}^\textrm{T}
\end{equation}
is the OFDM-IM subblock $g$ for transmit antenna $t$, and the corresponding channel and noise vectors are given respectively as
\begin{equation}
\mathbf{\breve{h }}_{r,t}^g = \big[ \begin{matrix}
\breve{h}_{r,t}^g(1) & \breve{h}_{r,t}^g(2) &\cdots & \breve{h}_{r,t}^g(N)
\end{matrix} \big]^\textrm{T}
\end{equation}
and 
\begin{equation}
\mathbf{\breve{w}}_{r}^g = \begin{bmatrix}
\breve{w}_{r}^g(1) & \breve{w}_{r}^g(2) &\cdots & \breve{w}_{r}^g(N)
\end{bmatrix}^\textrm{T}.
\end{equation}

For the presentation and analysis of different type of detectors, the following signal model is obtained from (\ref{eq:3}) for subcarrier $n$ of subblock $g$:
\begin{align}
\!\begin{bmatrix}
y_1^g(n) \\ y_2^g(n) \\ \vdots \\ y_R^g(n) \\ 
\end{bmatrix} \!\!\!&=\!\! \!
\begin{bmatrix}
\breve{h}_{1,1}^g(n)  & \cdots & \breve{h}_{1,T}^g(n) \\
\breve{h}_{2,1}^g(n) &  \cdots & \breve{h}_{2,T}^g(n) \\
\vdots  & \ddots & \vdots \\
\breve{h}_{R,1}^g(n) &  \cdots & \breve{h}_{R,T}^g(n)
\end{bmatrix} \!\!\!\!
\begin{bmatrix}
x_1^g(n) \\ x_2^g(n) \\ \vdots \\ x_T^g(n) \\ 
\end{bmatrix} \!\!\!+\!\!\! \begin{bmatrix}
\!\breve{w}_1^g(n)\! \\ \!\breve{w}_2^g(n)\! \\ \!\vdots\! \\ \!\breve{w}_R^g(n)\! \\ 
\end{bmatrix} \nonumber \\
&\hspace*{1.4cm}\mathbf{\bar{y}}_n^g=\mathbf{H}_n^g \mathbf{\bar{x}}_n^g + \mathbf{\bar{w}}_n^g
\label{eq:MMSE}
\end{align}
for $n=1,2,\ldots,N$ and $g=1,2,\ldots,G$, where $ \mathbf{\bar{y}}_n^g $ is the received signal vector, $ \mathbf{H}_n^g \in \mathbb{C}^{R\times T} $ is the corresponding channel matrix which contains the channel coefficients between transmit and receive antennas and assumed to be perfectly known at the receiver, $ \mathbf{\bar{x}}_n^g  $ is the data vector which contains the simultaneously transmitted symbols from all transmit antennas and can have zero terms due to index selection in each branch of the transmitter and $\mathbf{\bar{w}}_n^g$ is the noise vector.

The signal-to-noise ratio (SNR) is defined as $\mathrm{SNR}=E_b/N_{0,T}$ where $E_b=(N_F+C_p)/m$ [joules/bit] is the average transmitted energy per bit. The spectral efficiency of the MIMO-OFDM-IM scheme is calculated as $mT/(N_F+C_p)$ [bits/s/Hz].  


\section{ML Detection of MIMO-OFDM-IM}

In this section, we propose ML and near-ML detectors of the MIMO-OFDM-IM scheme which can be used in applications where the BER is critical. We also derive the ABEP of the brute-force ML detector which can be considered as a performance benchmark for the near-ML detector, whose theoretical performance analysis is intractable.

\subsection{Brute-Force ML Detecion of MIMO-OFDM-IM}

A straightforward solution to the detection problem of (\ref{eq:3}) is the use of ML detector which can be realized for each subblock $g$ as 
\begin{equation}
\left( \mathbf{x}_1^g,\ldots,\mathbf{x}_T^g\right)_{\textrm{ML}} \! = \arg \!\!\min_{\left( \mathbf{x}_1^g,...,\mathbf{x}_T^g\right) } \sum_{r=1}^{R}  \bigg\| \mathbf{y}_r^g -\sum_{t=1}^{T}\mbox{diag} \left( \mathbf{x}_t^g\right)  \mathbf{\breve{h }}_{r,t}^g\bigg\|^2.
\label{eq:ML} 
\end{equation} 
As seen from (\ref{eq:ML}), the ML detector has to make a joint search over all transmit antennas due the interference between the subblocks of different transmit antennas. 


In this subsection, the ABEP of MIMO-OFDM-IM is derived by the PEP calculation for MIMO-OFDM-IM subblocks. Since the pairwise error (PE) events within different subblocks are identical, it is sufficient to investigate the PE events within a single subblock to determine the overall system performance. 

Stacking the received signals in (\ref{eq:MMSE}) for $N$ consecutive subcarriers of a given subblock $g$, we obtain
\begin{align}
&\begin{bmatrix}
\mathbf{\bar{y}}_1^g \\ \mathbf{\bar{y}}_2^g \\ \vdots \\ \mathbf{\bar{y}}_N^g
\end{bmatrix}= \begin{bmatrix}
\mathbf{H}_1^g & \mathbf{0} & \ldots & \mathbf{0} \\
\mathbf{0} & \mathbf{H}_2^g & \ldots & \mathbf{0} \\
\vdots & & \ddots & \vdots \\
\mathbf{0} & \mathbf{0} & \ldots & \mathbf{H}_N^g
\end{bmatrix} \begin{bmatrix}
\mathbf{\bar{x}}_1^g \\ \mathbf{\bar{x}}_2^g \\ \vdots \\ \mathbf{\bar{x}}_N^g
\end{bmatrix}  + \begin{bmatrix}
\mathbf{\bar{w}}_1^g \\ \mathbf{\bar{w}}_2^g \\ \vdots \\ \mathbf{\bar{w}}_N^g
\end{bmatrix} \nonumber \\
& \hspace{2.3cm} \mathbf{y}^g=\mathbf{H}^g \mathbf{x}^g + \mathbf{w}^g
\label{ML_matrix}
\end{align} 
where $\mathbf{0}$ denotes the all-zero matrix with dimensions $R\times T$,  $\mathbf{y}^g \in \mathbb{C}^{RN\times 1}$ is the vector of stacked received signals for the corresponding subblock, $\mathbf{H}^g \in \mathbb{C}^{RN\times TN}$ is the block-diagonal channel matrix, $\mathbf{x}^g \in \mathbb{C}^{TN\times 1}$ is the equivalent data vector which has $(CM^K)^T$ possible realizations according to index modulation and $\mathbf{w}^g \in \mathbb{C}^{RN\times 1}$ is the noise vector. Using the matrix form given in (\ref{ML_matrix}), the ML detection of MIMO-OFDM-IM for each subblock $g$ can also be performed as
 \begin{equation}
  \left( \mathbf{x}^g\right)_{\textrm{ML}}  =  \arg\min_{\mathbf{x}^g}   \big\| \mathbf{y}^g - \mathbf{H}^g \mathbf{x}^g\big\|^2.
 \label{eq:ML2} 
 \end{equation}     
Considering the signal model of (\ref{ML_matrix}), for a given channel matrix $\mathbf{H}^g$, if $ \mathbf{x}^g $ is transmitted and it is erroneously detected as $ \mathbf{e}^g $, where
\begin{equation}
\mathbf{e}^g= \begin{bmatrix}
(\mathbf{\bar{e}}_1^g)^{\textrm{T}} & (\mathbf{\bar{e}}_2^g)^{\textrm{T}} & \cdots & (\mathbf{\bar{e}}_N^g)^{\textrm{T}} 
\end{bmatrix}^{\textrm{T}},
\end{equation}
the conditional PEP (CPEP) can be calculated as
\begin{equation}
P\left( \mathbf{x}^g \rightarrow \mathbf{e}^g \left. \right| \mathbf{H}^g \right) = P \left( \big\| \mathbf{y}^g - \mathbf{H}^g\mathbf{x}^g\big\|^2 > \big\| \mathbf{y}^g - \mathbf{H}^g \mathbf{e}^g\big\|^2  \right). 
\end{equation} 
After some algebra, the CPEP of the MIMO-OFDM-IM scheme is obtained as
\begin{align}
P\left( \mathbf{x}^g \rightarrow \mathbf{e}^g \left. \right| \mathbf{H}^g \right) &= P \left( \big\| \mathbf{H}^g  \mathbf{x}^g   \big\|^2 - \big\| \mathbf{H}^g  \mathbf{e}^g   \big\|^2 - 2 \Re\left\lbrace (\mathbf{y}^g)^\textrm{H}  \mathbf{H}^g \left( \mathbf{x}^g - \mathbf{e}^g \right)   \right\rbrace  >  0   \right) \nonumber \\  
&= P \left( - \big\| \mathbf{H}^g \left( \mathbf{x}^g - \mathbf{e}^g \right)   \big\|^2 - 2 \Re\left\lbrace (\mathbf{w}^g)^\textrm{H} \mathbf{H}^g \left( \mathbf{x}^g - \mathbf{e}^g \right)    \right\rbrace > 0   \right) \nonumber \\
& = P \left(D>0 \right)  
\end{align} 
where $D \sim \mathcal{N} \left(m_D,\sigma_D^2 \right) $ with  
\begin{align}
m_D &=- \big\| \mathbf{H}^g \left( \mathbf{x}^g - \mathbf{e}^g \right)   \big\|^2 \nonumber \\
\sigma_D^2 &= 2 N_{0,F}  \big\| \mathbf{H}^g \left( \mathbf{x}^g - \mathbf{e}^g \right)   \big\|^2,
\end{align}
for which we obtain
\begin{equation}
P\left( \mathbf{x}^g \rightarrow \mathbf{e}^g \left. \right| \mathbf{H}^g \right)= Q\left(\sqrt{\frac{\big\| \mathbf{H}^g \left( \mathbf{x}^g - \mathbf{e}^g \right)   \big\|^2}{2 N_{0,F}}} \right). 
\label{eq:CPEP}
\end{equation}
Using the alternative form of the Q-function \cite{Simon}, (\ref{eq:CPEP}) can be rewritten as
\begin{equation}
P\left( \mathbf{x}^g \rightarrow \mathbf{e}^g \left. \right| \mathbf{H}^g \right)= \frac{1}{\pi} \int_{0}^{\pi/2} \!\!\exp \left( - \frac{\big\| \mathbf{H}^g \left( \mathbf{x}^g - \mathbf{e}^g \right)   \big\|^2}{4 N_{0,F} \sin^2\theta} \right) d\theta.
\label{eq:CPEP2}
\end{equation} 
Integrating the CPEP in (\ref{eq:CPEP2}) over the probability density function (pdf) of $\Gamma = \big\| \mathbf{H}^g \left( \mathbf{x}^g - \mathbf{e}^g \right)   \big\|^2 $, the unconditional PEP (UPEP) of the MIMO-OFDM-IM scheme is obtained as
\begin{equation}
P\left( \mathbf{x}^g \rightarrow \mathbf{e}^g \right)= \frac{1}{\pi} \int_{0}^{\pi/2}  M_{\Gamma}\left(-\frac{1}{4 N_{0,F} \sin^2\theta}  \right) d\theta
\label{eq:UPEP}
\end{equation}  
where $M_{\Gamma}(t)$ is the moment generating function (mgf) of $\Gamma$. Expressing $\Gamma$ in quadratic form as
\begin{equation}
\Gamma= \sum_{n=1}^{N} \big\| \mathbf{H}_n^g \left( \mathbf{\bar{x}}_n^g - \mathbf{\bar{e}}_n^g \right)   \big\|^2 = \sum_{n=1}^{N} \sum_{r=1}^{R}  \left( \mathbf{H}_{n}^g\right)_{r*}  \mathbf{Q}_n^g \left( \mathbf{H}_{n}^g\right)_{r*}^\textrm{H}
\label{eq:Quad}
\end{equation}
where 
\begin{equation}
\mathbf{Q}_n^g=\left( \mathbf{\bar{x}}_n^g - \mathbf{\bar{e}}_n^g \right) \left( \mathbf{\bar{x}}_n^g - \mathbf{\bar{e}}_n^g \right)^\textrm{H}.
\end{equation}
According to the quadratic form of $\Gamma$ given in (\ref{eq:Quad}), its mgf is obtained as \cite{Turin}
\begin{equation}
M_{\Gamma}(t) \!=\! \prod\nolimits_{n=1}^{N} \! \left[ \det\left( \mathbf{I}_T \!-\! t \mathbf{L} \mathbf{Q}_n^g\right)  \right]^{-R} \! = \! \prod\nolimits_{n=1}^{N} \!\! \left(1\!-\!t  \big\|  \mathbf{\bar{x}}_n^g - \mathbf{\bar{e}}_n^g   \big\|^2  \right)^{\!\!-R} 
\end{equation}
since $\left( \mathbf{H}_{n}^g\right)_{r*} $'s are i.i.d. for all $r$ and $n$, we obtain
\begin{equation}
\mathbf{L}=E\left\lbrace \left( \mathbf{H}_{n}^g\right)_{r*}^\textrm{H} \left( \mathbf{H}_{n}^g\right)_{r*} \right\rbrace = \mathbf{I}_T 
\end{equation}
and $\mbox{rank}(\mathbf{Q}_n^g)=1$. Finally, from (\ref{eq:UPEP}), the UPEP of the MIMO-OFDM-IM scheme is obtained as 
\begin{equation}
P\left( \mathbf{x}^g \rightarrow \mathbf{e}^g \right)= \frac{1}{\pi} \int_{0}^{\pi/2}  \prod_{n=1}^{N} \Bigg( \frac{\sin^2\theta}{\sin^2\theta + \dfrac{\big\|  \mathbf{\bar{x}}_n^g - \mathbf{\bar{e}}_n^g   \big\|^2}{4 N_{0,F}}}\Bigg)^{R} d\theta. 
\label{eq:UPEP_final}
\end{equation}  
Please note that the integral given in (\ref{eq:UPEP_final}) has closed form solutions in Appendix 5A of \cite{Simon} for different $N$ values. \\ 
\textit{Remark 1}: For the worst case PE events in which there are no active indices errors and a single $M$-ary symbol is erroneously detected in $\mathbf{e}^g$, we obtain $ \big\|  \mathbf{\bar{x}}_n^g - \mathbf{\bar{e}}_n^g   \big\|^2 \neq 0  $ for only a single $n$ value. In this case, the diversity order of the MIMO-OFDM-IM scheme is calculated after some manipulation as\cite{Jafarkhani}
\begin{equation}
G_d =- \lim\limits_{N_{0,F} \rightarrow 0} \frac{\log \left( P\left( \mathbf{x}^g \rightarrow \mathbf{e}^g \right) \right) }{\log \left( 1/ N_{0,F} \right) } = R.
\end{equation}
On the other hand, the distance spectrum of the MIMO-OFDM-IM is improved due to the PE events in which there are errors in active indices since these PE events have lower occurrence probabilities.\\
\textit{Remark 2}: After the evaluation of the UPEP, the ABEP of the MIMO-OFDM-IM scheme can be obtained by the asymptotically tight union upper bound as
\begin{equation}
P_b \le \frac{1}{n_{b} n({\mathbf{x}^g})} \sum\nolimits_{\mathbf{x}^g}\sum\nolimits_{\mathbf{e}^g}P\left( \mathbf{x}^g \rightarrow \mathbf{e}^g \right) n(\mathbf{x}^g ,\mathbf{e}^g ) 
\label{eq:16}
\end{equation}
where  $n_{b}=pT$ is the total number of information bits carried by $\mathbf{x}^g$, $n({\mathbf{x}^g})=(CM^K)^T$ is the total number of possible realizations of $\mathbf{x}^g$ and $n(\mathbf{x}^g ,\mathbf{e}^g )$ is the number of bit errors for the corresponding PE event $(\mathbf{x}^g \rightarrow \mathbf{e}^g)$.

\subsection{Simplified Near-ML Detection of MIMO-OFDM-IM}
The total decoding complexity of the brute-force ML detector given in (\ref{eq:ML}) and (\ref{eq:ML2}) in terms of complex multiplications (CMs) is $ \sim \mathcal{O}(M^{KT})$, which is considerably higher than that of classical MIMO-OFDM, whose complexity is $ \sim \mathcal{O} (M^{T})$, even if assuming that both schemes use the same constellation order. In this section, we propose a near-ML detector for the MIMO-OFDM-IM scheme which has the same order decoding complexity compared to classical MIMO-OFDM ML detector.


The ML detector in (\ref{eq:ML}) maximizes the joint conditional pdf of $p\left( \mathbf{y}_1^g,\ldots,\mathbf{y}_R^g \left. \right| \mathbf{x}_1^g,\ldots,\mathbf{x}_T^g \right) $, i.e., jointly searches for $\mathbf{x}_1^g,\ldots,\mathbf{x}_T^g$ using the reference look-up table. On the other hand, the proposed near-ML detector calculates a probabilistic measure for each element $(\mathbf{x}_t^g)$ of the reference look-up table for a given transmit antenna; therefore, it reduces the size of the search space considerably. For this purpose, the near-ML detector considers the model in (\ref{eq:MMSE}) and implements the following steps:\footnote{For notational simplicity, the realizations of the random vectors/variables are dropped in (\ref{eq:18})-(\ref{eq:20}). See Example 1 for more details.}
\begin{enumerate}
\item Calculate $N(M+1)^T$ different conditional probability values of  $P\left(\mathbf{\bar{x}}_n^g \left. \right| \mathbf{\bar{y}}_n^g  \right)$ by considering 
\begin{equation}
\hspace*{-0.6cm}P\left(\mathbf{\bar{x}}_n^g \left. \right| \mathbf{\bar{y}}_n^g  \right)  = \frac{f\left(\mathbf{\bar{y}}_n^g \left. \right| \mathbf{\bar{x}}_n^g  \right) P\left(\mathbf{\bar{x}}_n^g \right)  }{f\left(\mathbf{\bar{y}}_n^g \right)}= \frac{f\left(\mathbf{\bar{y}}_n^g \left. \right| \mathbf{\bar{x}}_n^g  \right) P\left(\mathbf{\bar{x}}_n^g \right)  }{\displaystyle\sum_{\mathbf{\bar{x}}_n^g}^{}f\left(\mathbf{\bar{y}}_n^g \left. \right| \mathbf{\bar{x}}_n^g  \right) P\left(\mathbf{\bar{x}}_n^g \right)}
\label{eq:18}
\end{equation} 
for $n=1,\ldots,N$ and all possible $\mathbf{\bar{x}}_n^g$ vectors where conditioned on $\mathbf{\bar{x}}_n^g $, $ \mathbf{\bar{y}}_n^g   $ has the multivariate complex Gaussian distribution with pdf
\begin{equation}
f\left(\mathbf{\bar{y}}_n^g \left. \right| \mathbf{\bar{x}}_n^g  \right)= \left( \pi N_{0,F}\right)^{-R}    \exp\left( - \frac{\left\| \mathbf{\bar{y}}_n^g-\mathbf{H}_n^g \mathbf{\bar{x}}_n^g \right\|^2}{N_{0,F}}  \right).
\end{equation}
\item In order to obtain the occurrence probability for each element $(\mathbf{x}_t^g)$ of the reference look-up table, calculate
\begin{equation}
\hspace*{-0.6cm}P\left(\mathbf{x}_t^g \right)= \prod_{n=1}^{N} P\left(x_t^g(n) \right) = \prod_{n=1}^{N} \sum_{\mathbf{\bar{x}}_n^g, \bar{x}_n^g(t)=x_t^g(n) }^{} P\left(\mathbf{\bar{x}}_n^g \left. \right| \mathbf{\bar{y}}_n^g  \right)
\label{eq:19} 
\end{equation}         
where $\bar{x}_n^g(t)$ is $t$th element of $\mathbf{\bar{x}}_n^g \in \mathbb{C}^{T \times 1}$. (\ref{eq:19}) provides a clever way to transform the probabilities of $P\left(\mathbf{\bar{x}}_n^g \left. \right| \mathbf{\bar{y}}_n^g  \right)$, which consider the symbols transmitted from different antennas, into the probabilities of $P\left(\mathbf{x}_t^g \right)$, which are distinct for each transmit antenna.
\item Finally, after the calculation of $CM^K$ probability values for each transmit antenna $t$, decide on the most likely element of the reference look-up table as
\begin{equation}
(\mathbf{x}_t^g)_{\textrm{near-ML}} = \arg \max\nolimits_{\mathbf{x}_t^g}  P\left(\mathbf{x}_t^g \right). 
\label{eq:20}
\end{equation}       
\end{enumerate}

As seen from (\ref{eq:18})-(\ref{eq:19}), for the calculation of $ P\left(\mathbf{x}_t^g \right) $ values, a search over the possible realizations of $  \mathbf{\bar{x}}_n^g $ has to be made, which reduces the size of the search space to $\left( M+1 \right)^T $ for each $n$ value since $\bar{x}_n^g(t) \in \left\lbrace 0, \mathcal{S} \right\rbrace $ and a total of $\sim \mathcal{O}(M^T)$ CMs are required. The following numerical example shows the steps of the near-ML detector. \\
\textit{Example 1}: Consider the MIMO-OFDM-IM scheme with the following system parameters: $T=M=K=2$, $N=4$. For these values, the reference look-up table contains $CM^K=16$ elements. In this case, $(\mathbf{\bar{x}}_n^g)^\text{T} $ has $(M+1)^{T}=9$ possible realizations: $\begin{bmatrix}
0 & 0
\end{bmatrix}$, $\begin{bmatrix}
0 & 1
\end{bmatrix}$,$\begin{bmatrix}
0 & -1
\end{bmatrix}$, $\begin{bmatrix}
1 & 0
\end{bmatrix}$, $\begin{bmatrix}
1 & 1
\end{bmatrix}$, $\begin{bmatrix}
1 & -1
\end{bmatrix}$, $\begin{bmatrix}
-1 & 0
\end{bmatrix}$, $\begin{bmatrix}
-1 & 1
\end{bmatrix}$ and $\begin{bmatrix}
-1 & -1
\end{bmatrix}$, with the following probabilities: $0.25,0.125,0.125$, $0.125$,$0.0625,0.0625,0.125$, $0.0625$ and $0.0625$, respectively. 

First, the near-ML detector calculates and stores the probabilities $ P\left(\mathbf{\bar{x}}_n^g \left. \right| \mathbf{\bar{y}}_n^g  \right)  $ using the received signals $\mathbf{\bar{y}}_n^g$ and possible $ \mathbf{\bar{x}}_n^g $ vectors for $n=1,2,3,4$, where a total of $N(M+1)^T=36$ probability calculations are required. As an example, for $n=1$, nine probability values of $ P\left(\mathbf{\bar{x}}_1^g \left. \right| \mathbf{\bar{y}}_1^g  \right)  $  are calculated and stored using (\ref{eq:18}).

Second, the occurrence probability of the each element of the reference look-up table is calculated from (\ref{eq:19}). Let us consider that we want to calculate the probability of $P(\mathbf{x}_1^g=\begin{bmatrix}
1 & 0 & -1 & 0
\end{bmatrix}^\text{T})$, where $ \begin{bmatrix}
1 & 0 & -1 & 0
\end{bmatrix}^\text{T} $ is selected from the look-up table (Table I). According to (\ref{eq:19}), we have
\begin{align*}
& P(\mathbf{x}_1^g=\begin{bmatrix}
1 & 0 & -1 & 0
\end{bmatrix}^\text{T}) =P(x_1^g(1)=1) P(x_1^g(2)=0) P(x_1^g(3)=-1) P(x_1^g(4)=0)  
\end{align*}
where 
\begin{align*}
 P(x_1^g(1)=1)& = \sum_{\mathbf{\bar{x}}_1^g,\bar{x}_1^g(1)=1}^{} P\left(\mathbf{\bar{x}}_1^g \left. \right| \mathbf{\bar{y}}_1^g  \right) \nonumber \\
& =P\left( \mathbf{\bar{x}}_1^g=\begin{bmatrix}
1 & 0
\end{bmatrix}^\text{T} \left.  \right| \mathbf{\bar{y}}_1^g\right)  + P\left( \mathbf{\bar{x}}_1^g=\begin{bmatrix}
1 & 1
\end{bmatrix}^\text{T} \left.  \right| \mathbf{\bar{y}}_1^g\right) + P\left( \mathbf{\bar{x}}_1^g=\begin{bmatrix}
1 & -1
\end{bmatrix}^\text{T} \left.  \right| \mathbf{\bar{y}}_1^g\right),    \\
 P(x_1^g(2)=0) & = \sum_{\mathbf{\bar{x}}_2^g,\bar{x}_2^g(1)=0}^{}  P\left(\mathbf{\bar{x}}_2^g \left. \right| \mathbf{\bar{y}}_2^g  \right) \nonumber \\
 &=P\left( \mathbf{\bar{x}}_2^g=\begin{bmatrix}
0 & 0
\end{bmatrix}^\text{T}  \left.  \right| \mathbf{\bar{y}}_2^g\right)  + P\left( \mathbf{\bar{x}}_2^g=\begin{bmatrix}
0 & 1
\end{bmatrix}^\text{T} \left.  \right| \mathbf{\bar{y}}_2^g\right) + P\left( \mathbf{\bar{x}}_2^g=\begin{bmatrix}
0 & -1
\end{bmatrix}^\text{T} \left.  \right| \mathbf{\bar{y}}_2^g\right),   \\
 P(x_1^g(3)=-1) &= \sum_{\mathbf{\bar{x}}_3^g,\bar{x}_3^g(1)=-1}^{}  P\left(\mathbf{\bar{x}}_3^g \left. \right| \mathbf{\bar{y}}_3^g  \right)  \nonumber \\
 &=  P\left( \mathbf{\bar{x}}_3^g=  \begin{bmatrix}
-1 & 0
\end{bmatrix}^\text{T}\left.  \right| \mathbf{\bar{y}}_3^g\right)  + P\left( \mathbf{\bar{x}}_3^g=\begin{bmatrix}
-1 & 1
\end{bmatrix}^\text{T} \left.  \right| \mathbf{\bar{y}}_3^g\right) + P\left( \mathbf{\bar{x}}_3^g=\begin{bmatrix}
-1 & -1
\end{bmatrix}^\text{T} \left.  \right| \mathbf{\bar{y}}_3^g\right),    \\
 P(x_1^g(4)=0)&= \sum_{\mathbf{\bar{x}}_4^g,\bar{x}_4^g(1)=0}^{}P\left(\mathbf{\bar{x}}_4^g \left. \right| \mathbf{\bar{y}}_4^g  \right) \nonumber \\
 &=P\left( \mathbf{\bar{x}}_4^g=\begin{bmatrix}
0 & 0
\end{bmatrix}^\text{T} \left.  \right| \mathbf{\bar{y}}_4^g\right)  + P\left( \mathbf{\bar{x}}_4^g=\begin{bmatrix}
0 & 1
\end{bmatrix}^\text{T} \left.  \right| \mathbf{\bar{y}}_4^g\right) + P\left( \mathbf{\bar{x}}_4^g=\begin{bmatrix}
0 & -1
\end{bmatrix}^\text{T} \left.  \right| \mathbf{\bar{y}}_4^g\right).   
\end{align*} 
Similarly, $P(\mathbf{x}_2^g=\begin{bmatrix}
1 & 0 & -1 & 0
\end{bmatrix}^\text{T})$ can be calculated considering the second elements of $ \mathbf{\bar{x}}_n^g $. Finally, the most likely element of the look-up table is determined from (\ref{eq:20}) after the calculation of $CM^K$ probability values, where $\sum_{\mathbf{x}_t^g}^{}P(\mathbf{x}_t^g)=1$.

\section{MMSE Detection of MIMO-OFDM-IM Scheme}

The total decoding complexity of the brute-force and near-ML detectors can be considerably high for higher order modulations and MIMO systems. In this section, instead of the exponentially increasing decoding complexity of the ML detectors, we deal with the MMSE detection of the MIMO-OFDM-IM scheme, which significantly reduces the decoding complexity. The approximate ABEP of the newly proposed simple MMSE detector is also provided which can be considered as a reference for MMSE and LLR detector.

\subsection{Simple MMSE Detection of MIMO-OFDM-IM}
 Let us reconsider the following signal model which is obtained from (\ref{eq:3}) for subcarrier $n$ of subblock $g$:
\begin{equation}
\mathbf{\bar{y}}_n^g=\mathbf{H}_n^g \mathbf{\bar{x}}_n^g + \mathbf{\bar{w}}_n^g
\label{eq:MMSE2}
\end{equation} 
for $n=1,2,\ldots,N$ and $g=1,2,\ldots,G$.
For classical MIMO-OFDM, the data symbols can be simply recovered after processing the received signal vector in (\ref{eq:MMSE2}) with the MMSE detector. On the other hand, due to the index information carried by the subblocks of the proposed scheme, it is not possible to detect the transmitted symbols by only processing $ \mathbf{\bar{y}}_n^g $ for a given subcarrier $n$ in the MIMO-OFDM-IM scheme. Therefore, $N$ independent and successive MMSE detections are performed for the proposed scheme using the MMSE filtering matrix \cite{MMSE}
\begin{equation}
\mathbf{W}_{n}^{g} = \left( \left( \mathbf{H}_n^g\right)^\text{H}  \mathbf{H}_n^g + \frac{\mathbf{I}_T}{\rho}  \right)^{-1} \left( \mathbf{H}_n^g\right)^\text{H}
\end{equation}
for $n=1,2,\ldots,N$, where $\rho=\sigma_x^2 / N_{0,F}$, $\sigma_x^2=K/N$ and
\begin{equation}
E\big\lbrace \mathbf{\bar{x}}_n^g \left( \mathbf{\bar{x}}_n^g\right)^\text{H} \big\rbrace = \sigma_x^2 \mathbf{I}_T
\end{equation}
due to zero terms in $\mathbf{\bar{x}}_n^g$ come from index selection. By the left multiplication of $ \mathbf{\bar{y}}_n^g $ given in (\ref{eq:MMSE2}) with $ \mathbf{W}_n^g $, MMSE detection is performed as
\begin{equation}
\mathbf{z}_n^g=\mathbf{W}_n^g \mathbf{\bar{y}}_n^g=\mathbf{W}_n^g \mathbf{H}_n^g \mathbf{\bar{x}}_n^g + \mathbf{W}_n^g \mathbf{\bar{w}}_n^g
\end{equation}    
where 
\begin{equation}
\mathbf{z}_n^g=\begin{bmatrix}
z_n^g(1) & z_n^g(2) & \cdots & z_n^g(T)
\end{bmatrix}^\text{T}
\end{equation}
is the MMSE estimate of $ \mathbf{\bar{x}}_n^g $. The MMSE estimate of MIMO-OFDM-IM subblocks
\begin{equation}
\mathbf{\hat{x}}_t^g=\begin{bmatrix}
\hat{x}_t^g(1) & \hat{x}_t^g(2)  & \cdots & \hat{x}_t^g(N) 
\end{bmatrix}^\text{T}
\end{equation}
can be obtained by rearranging the elements of $\mathbf{z}_n^g,n=1,2,\ldots,N$ as
\begin{equation}
\mathbf{\hat{x}}_t^g=\begin{bmatrix}
z_1^g(t) & z_2^g(t)  & \cdots & z_N^g(t) 
\end{bmatrix}^\text{T}
\end{equation}
for $t=1,2,\ldots,T$ and $g=1,2,\ldots,G$. 

After MMSE filtering and rearranging the elements of MMSE filtered signals, the interference between the subblocks of different transmit antennas is eliminated and the transmitted subblocks can be determined by considering the conditional multivariate pdf of $\mathbf{\hat{x}}_t^g$ given as \cite{Complex_Gauss}
\begin{equation}
f\left( \mathbf{\hat{x}}_t^g \left. \right| \mathbf{x}_t^g \right)=\frac{\pi^{-N}}{\det(\mathbf{ C})} \exp\left( - \left( \mathbf{\hat{x}}_t^g- \mathbf{m}   \right)^\text{H} \mathbf{C}^{-1} \left( \mathbf{\hat{x}}_t^g- \mathbf{m}   \right)  \right)  
\end{equation}
where considering that $N$-successive MMSE operations are independent, the conditional mean and covariance matrix of $\mathbf{\hat{x}}_t^g$ conditioned on $\mathbf{x}_t^g$ are calculated respectively as
\begin{align}
\mathbf{m} & =E\left\lbrace \mathbf{\hat{x}}_t^g \right\rbrace = \begin{bmatrix}
 Q_1 x_t^g(1) & \cdots & \  Q_Nx_t^g(N) 
\end{bmatrix}^\text{T}  \nonumber \\
\mathbf{C} & = \mbox{cov}\left( \mathbf{\hat{x}}_t^g  \right)= \mbox{diag} \left( \begin{bmatrix}
C_1 & \cdots & C_N
\end{bmatrix} \right). 
\label{eq:26} 
\end{align}
For notational simplicity we defined
\begin{align}
Q_n &\triangleq (\mathbf{W}_n^g \mathbf{H}_n^g)_{t,t}, \nonumber \\
C_n &\triangleq (\mbox{cov}(\mathbf{z}_n^g))_{t,t}
\end{align}
for $n=1,2\ldots,N$ in (\ref{eq:26}). The statistics of $\mathbf{\hat{x}}_t^g$ are obtained from the conditional statistics of $\mathbf{z}_n^g$ conditioned on $x_t^g(n)\in\left\lbrace 0,\mathcal{S}\right\rbrace $, which are given in \cite{MIMO_OFDM_IM} as
\begin{align}
E\left\lbrace \mathbf{z}_n^g \right\rbrace &= \mathbf{W}_n^g \mathbf{H}_n^g E\left\lbrace  \mathbf{\bar{x}}_n^g  \right\rbrace = \left( \mathbf{W}_n^g \mathbf{H}_n^g\right)_{*t} x_t^g(n) \nonumber \\
\mbox{cov}(\mathbf{z}_n^g) &=  \mathbf{W}_n^g \mathbf{H}_n^g \mbox{cov}\left(\mathbf{\bar{x}}_n^g  \right) \left(\mathbf{H}_n^g \right)^\text{H}  \!\! \left(\mathbf{W}_n^g \right)^\text{H} + N_{0,F}\mathbf{W}_n^g \left(\mathbf{W}_n^g \right)^\text{H}  
\label{eq:cov}
\end{align}
where 
\begin{equation}
\mbox{cov}\left(\mathbf{\bar{x}}_n^g  \right) = \mbox{diag}\left( \begin{bmatrix}
\sigma_x^2 &  \ldots & \sigma_x^2 & 0 & \sigma_x^2 & \ldots & \sigma_x^2  
\end{bmatrix}  \right) 
\end{equation}
is a diagonal matrix whose $t$th diagonal element is zero.

The simple MMSE detector decides onto the most likely subblock by maximizing the conditional pdf of $\mathbf{\hat{x}}_t^g$ as
\begin{align}
\left( \mathbf{x}_t^g\right)_{\textrm{MMSE}} = \arg   \max_{\mathbf{x}_t^g}   f  \left( \mathbf{\hat{x}}_t^g \! \left.  \right| \mathbf{x}_t^g \right)  =  \arg \!\min_{\mathbf{x}_t^g}   \sum_{n=1}^{N}  \frac{\left| \hat{x}_t^g(n) - Q_n x_t^g(n) \right|^2 }{C_n}
\label{eq:MMSE_metrik}
\end{align}
where (\ref{eq:MMSE_metrik}) is obtained by dropping the constant terms and considering the diagonal structure of $\mathbf{C}$. The CPEP of erroneously deciding in favor of $\mathbf{e}_t^g=\begin{bmatrix}
e_t^g(1) & e_t^g(2) &\cdots &e_t^g(N)
\end{bmatrix}^\textrm{T}$ given that $\mathbf{x}_t^g$ is transmitted can be calculated as
 \begin{align}
 & P\left( \mathbf{x}_t^g \rightarrow  \mathbf{e}_t^g \left. \right| \mathbf{H}_1^g, \cdots, \mathbf{H}_N^g  \right) \nonumber \\
 &=  P \left( \sum_{n=1}^{N} \frac{\left| \hat{x}_t^g(n)-Q_n x_t^g(n) \right|^2 - \left| \hat{x}_t^g(n)-Q_n e_t^g(n) \right|^2  }{C_n}  >0   \right)  	\nonumber \\
 &= P \Bigg( \sum_{n=1}^{N}  \Big( Q_n^2\big( \left|x_t^g(n) \right|^2- \left|e_t^g(n) \right|^2  \big) - 2 \Re\big\lbrace Q_n \hat{x}_t^g(n)    \left( x_t^g(n)- e_t^g(n) \right)^*   \big\rbrace \Big)   /  C_n > 0 \Bigg)  =  P \left(  D>0 \right)  
 \end{align}
where $D \sim \mathcal{N}\left(m_D,\sigma_D^2 \right) $ and considering 
\begin{align}
E\left\lbrace \hat{x}_t^g(n) \right\rbrace&=Q_n x_t^g(n) \nonumber \\
Var \left(\hat{x}_t^g(n) \right)&=C_n
\end{align}
and $ Var(\Re \left\lbrace \hat{x}_t^g(n)  \right\rbrace  ) = Var(\Im \left\lbrace \hat{x}_t^g(n)  \right\rbrace  )=C_n/2 $ for complex $M$-QAM constellation symbols with i.i.d. real and imaginary parts, we obtain 
\begin{align}
m_D  &= - \sum\nolimits_{n=1}^{N} V_n \Delta_n, \nonumber \\
\sigma_D^2 &= 2\sum\nolimits_{n=1}^{N} V_n \Delta_n
\end{align}
 which yields the following CPEP 
\begin{equation}
P\left( \mathbf{x}_t^g \rightarrow  \mathbf{e}_t^g \left. \right| \mathbf{H}_1^g, \cdots, \mathbf{H}_N^g  \right) = Q \left( \sqrt{\sum\nolimits_{n=1}^{N} V_n \Delta_n }\right) 
\label{CPEP_32}
\end{equation}
where $V_n\triangleq\dfrac{Q_n^2}{2C_n}$ and $\Delta_n\triangleq\left|x_t^g(n)- e_t^g(n)  \right|^2$. The r.v. $V_n$, which is the ratio of two correlated r.v.'s, has a nonparametric pdf which is a function of the SNR and is the same for all $n$ and $t$. Therefore, we provide an upper bound for the UPEP as follows. 

Assuming  $Q_n^2 \approx 1$ and 
\begin{equation}
C_n \approx N_{0,F} \left( (\mathbf{H}_n^g)^{-1} ((\mathbf{H}_n^g)^{-1})^\text{H} \right)_{t,t}
\end{equation}
for $N_{0,F} \ll 1$, which is quite reasonable for practical applications, and using alternative form of the Q-function, we obtain
\begin{equation}
P\left( \mathbf{x}_t^g \rightarrow  \mathbf{e}_t^g \left. \right| \mathbf{H}_1^g, \cdots,\! \mathbf{H}_N^g  \right) < \frac{1}{\pi} \!\! \int\limits_{0}^{\pi/2} \! \! \exp\left( - \frac{ \sum\nolimits_{n=1}^N Z_n \Delta_n   }{4 N_{0,F} \sin^2 \theta} \right) d\theta
\label{CPEP_final}
\end{equation}
where 
\begin{equation}
Z_n \triangleq 1/  \left( \left[ (\mathbf{H}_n^g)^\text{H} \mathbf{H}_n^g  \right]^{-1}  \right)_{t,t}   
\end{equation}
and the inequality arises from $V_n\approx Z_n/(2N_{0,F}) < V_n $. Considering that $Z_n$ is exponentially distributed for $T=R$  \cite{MMSE} with mgf 
\begin{equation}
M_{Z_n}(t)=1/(1-t)
\end{equation}
for all $n$ and $t$, integrating (\ref{CPEP_final}) over the pdf's of $Z_n,n=1,2,\ldots,N$, the UPEP of the simple MMSE detector is obtained as 
\begin{align}
P\left( \mathbf{x}_t^g \rightarrow  \mathbf{e}_t^g \right) &< \frac{1}{\pi}\int\limits_{0}^{\pi/2} \prod\limits_{n=1}^{N} M_{Z_n}\left(\frac{-\Delta_n}{4 N_{0,F} \sin^2 \theta} \right) d\theta \nonumber \\
&= \frac{1}{\pi}\int\limits_{0}^{\pi/2} \prod\limits_{n=1}^{N} \left(  \frac{\sin^2 \theta}{\sin^2 \theta + \frac{\Delta_n}{4 N_{0,F}} } \right) d\theta
\label{UPEP}
\end{align}
which has a closed form solution in Appendix 5A of \cite{Simon}. \\
\textit{Remark 1}: The diversity order of the simple MMSE detector is equal to one considering the worst case PE events in (\ref{UPEP}). The UPEP in (\ref{UPEP}) is independent of the number of transmit antennas while $T=R$, and can be considered as an error performance upper bound for the MIMO-OFDM-IM scheme.\\
\textit{Remark 2}: To obtain a tighter UPEP approximation, the averaging over $V_n$ can be performed with a semi-analytical approach as follows. Considering the worst case PE events only in (\ref{CPEP_32}), where $\Delta_n \neq 0$ for a single $n \in \left\lbrace 1,2,\ldots,N \right\rbrace  $ value, we obtain
\begin{equation}
P\left( \mathbf{x}_t^g \rightarrow  \mathbf{e}_t^g \left. \right| \mathbf{H}_1^g, \cdots, \mathbf{H}_N^g  \right) \approx Q \left( \sqrt{ V_n \Delta_n }\right). 
\label{CPEP_worst}
\end{equation}
 For each SNR value, generating and storing $S=10^6$ samples of $V_n$ as $V_n(s),s=1,2,\ldots,S$ (as an example for $n=t=1$ since the distribution of $V_n$ is the same for all $ n $ and $ t $), the UPEP of the MIMO-OFDM-IM scheme can be obtained as
 \begin{equation}
 P\left( \mathbf{x}_t^g \rightarrow  \mathbf{e}_t^g \right) \approx \frac{1}{S} \sum\limits_{s=1}^{S}  Q \left( \sqrt{ V_n(s) \Delta_n }\right) 
 \label{eq:UPEP_semi}
 \end{equation}
\textit{Remark 3}: After the evaluation of the UPEP, ABEP of the simple MMSE detector can be obtained as
\begin{equation}
P_b \sim  \frac{1}{p n(\mathbf{x}_t^g)} \sum\limits_{\mathbf{x}_t^g}\sum\limits_{\mathbf{e}_t^g}P\left( \mathbf{x}_t^g \rightarrow \mathbf{e}_t^g \right) n(\mathbf{x}_t^g ,\mathbf{e}_t^g ) 
\end{equation}
where  $p$ is the total number of information bits carried by $\mathbf{x}_t^g$, $n(\mathbf{x}_t^g)=CM^K$ is the total number of possible realizations of $\mathbf{x}_t^g$ and $n(\mathbf{x}_t^g ,\mathbf{e}_t^g )$ is the number of bit errors for the corresponding PE event.

\subsection{MMSE and LLR Detection of MIMO-OFDM-IM}

MMSE and LLR detector of the MIMO-OFDM-IM scheme is proposed in \cite{MIMO_OFDM_IM} to implement a low complexity MMSE detection. Considering the conditional statistics of $\hat{x}_t^g(n)$, the MMSE and LLR detector of the MIMO-OFDM-IM scheme calculates the following LLR value for the $n$th subcarrier of $t$th transmitter for subblock $ g $ as
\begin{equation}
\lambda_t^g(n)=\ln \left(  \sum_{m=1}^{M}  \exp\left(  - \frac{\left|\hat{x}_t^g(n)  - Q_n s_{m}\right|^2 }{C_n}\right)  \right) + \frac{\left|\hat{x}_t^g(n) \right|^2 }{C_n}
\label{eq:15}
\end{equation}
for $n=1,2,\ldots,N$, $t=1,2,\ldots,T$ and $g=1,2,\ldots,G$, where $s_m \in \mathcal{S}$. For the case of reference look-up tables, the MMSE-LLR detector calculates LLR sums for each element of the look-up table and determines the active indices which provide the highest LLR sum. Details of this method can be found in \cite{MIMO_OFDM_IM}.

In case of the selection of active indices with combinatorial algorithm, after the calculation of $N$ LLR values form (\ref{eq:15}) for each subblock, the detector decides on $K$ active indices out of them having maximum LLR values. Denoting the indices of these subcarriers by 
\begin{equation}
\mathbf{\hat{j}}^g_t=\begin{bmatrix}
\hat{j}_t^g(1) & \hat{j}_t^g(2) &\cdots &\hat{j}_t^g(K)
\end{bmatrix}^\mathrm{T},
\end{equation}
the corresponding index selecting bits can be determined with demapping algorithm, and the $M$-ary symbols transmitted by the active subcarriers are determined for $k=1,2,\ldots,K$ as
\begin{equation}
(s_t^g(k))_{\text{MMSE}}=\arg\min_{s_m \in \mathcal{S}}	\Big|\hat{x}_t^g(\hat{j}_t^g(k))- Q_{\hat{j}_t^g(k)} s_m \Big|^2 
\label{eq:17}
\end{equation} 
where $Q_n$ is defined in (\ref{eq:26}). The above process is repeated for each subblock of each transmit antenna. As seen from (\ref{eq:15})- (\ref{eq:17}), the complexity of the MMSE-LLR detector, in terms of CMs, is $\sim \mathcal{O}(M)$ per subcarrier.

\subsection{MMSE and LLR Detection of MIMO-OFDM-IM with OSIC}
Successive interference cancellation (SIC) techniques have been efficiently implemented for MMSE detection based V-BLAST systems to obtain better BER performance with the price of increased decoding complexity \cite{MMSE_OSIC}. In other words, MMSE with SIC is an intermediate solution between ML and classical MMSE detections and provides a trade-off in performance and complexity. In this subsection, a novel OSIC based sequential MMSE-LLR detector is proposed for the MIMO-OFDM-IM scheme to further improve the error performance.   

The OSIC-MMSE-LLR detector of the MIMO-OFDM-IM scheme considers $N$ successive MMSE detections and obtains the following empirical min-max metric for each subblock $g$ of each transmit antenna $t,t=1,2,\ldots,T$ as
\begin{equation}
\gamma_t = \max \left\lbrace \Big\| \left( \left( \mathbf{G}_1^g\right)^{+} \right)_{t*}  \Big\|^2,\ldots, \Big\| \left( \left( \mathbf{G}_N^g\right)^{+} \right)_{t*}  \Big\|^2 \right\rbrace 	
\end{equation}
where 
\begin{equation}
\mathbf{G}_n^g=\begin{bmatrix}
\mathbf{H}_n^g \\ (1/\sqrt{\rho}) \mathbf{I}_{T}
\end{bmatrix}
\end{equation}
and 
\begin{equation}
\left( \mathbf{G}_n^g\right)^{+}=\left( \left( \mathbf{G}_n^g \right)^\text{H}  \mathbf{G}_n^g \right)^{-1} \left( \mathbf{G}_n^g \right)^\text{H} .
\end{equation}
The subblock of the transmit antenna with the minimum metric $(\gamma_t)$ is selected as the best subblock in terms of signal-to-interference-plus-noise ratio (SINR) and ordering is performed according to $\gamma_t,t=1,2,\ldots,T$. For each transmit antenna, the MMSE estimate of MIMO-OFDM-IM subblocks are obtained as $\hat{x}_t^g(n)=\mathbf{q}_{n,t}^g \mathbf{\bar{y}}_n^g$, where $\mathbf{q}_{n,t}^g \triangleq \left( \mathbf{W}_n^g \right) _{t*} \in \mathbb{C}^{1\times R}$ for $n=1,2,\ldots,N$. The conditional mean and variance $(\tilde{C}_n)$ of Gaussian distributed $ \hat{x}_t^g(n) $ is calculated for this case respectively as
\begin{align}
E\left\lbrace \hat{x}_t^g(n) \right\rbrace &= \left( \mathbf{q}_{n,t}^g \mathbf{H}_n^g \right)_{*t} x_t^g(n)= \tilde{Q}_n x_t^g(n) \nonumber \\
Var \left( \hat{x}_t^g(n)  \right) & = \mathbf{q}_{n,t}^g \mathbf{H}_n^g \mbox{cov}\left(\mathbf{\bar{x}}_n^g  \right)  \left(\mathbf{H}_n^g \right)^\text{H} \!\! \left( \mathbf{q}_{n,t}^g\right) ^\text{H} \!\!\!+\! N_{0,F} \! \left\| \mathbf{q}_{n,t}^g  \right\|^2 
\end{align}
 where $ \mbox{cov}\left(\mathbf{\bar{x}}_n^g  \right) \in \mathbb{R}^{T\times T}   $ is defined in (\ref{eq:cov}). Once the new statistics of $ \hat{x}_t^g(n) $ are obtained, the same procedures explained in Subsection IV-B are followed to determine the active indices and $M$-ary constellation symbols where the corresponding LLR values are calculated for this case as
\begin{equation}
\lambda_t^g(n)\!=\!\ln \!\Bigg(  \sum_{m=1}^{M} \! \exp\Bigg(  - \frac{\big|\hat{x}_t^g(n)  \!- \!\tilde{Q}_n s_{m}\big|^2 }{\tilde{C}_n}\Bigg)  \Bigg)\! + \frac{\left|\hat{x}_t^g(n) \right|^2 }{\tilde{C}_n}.
\label{eq:42}
\end{equation}  
According to the MMSE-LLR-OSIC detection, once the estimate $(\mathbf{x}_t^g)_{\text{MMSE}}$ is obtained, the received signal vectors are updated for $n=1,2,\ldots,N$ as  
\begin{equation}
\mathbf{\bar{y}}_n^g = \mathbf{\bar{y}}_n^g - \mathbf{H}_n^g (\mathbf{\bar{x}}_n^g)_{\text{MMSE}} 
\end{equation}
where
\begin{equation}
(\mathbf{\bar{x}}_n^g)_{\text{MMSE}}= \begin{bmatrix}
0 & \cdots & 0 & (x_t^g(n))_{\text{MMSE}} & 0 \cdots & 0
\end{bmatrix}^\text{T}
\end{equation}
is an all-zero vector except its $t$th element being $(x_t^g(n))_{\text{MMSE}}$ and $\left( \mathbf{H}_n^g\right)_{*t},n=1,2,\ldots,N$ are filled with zeros according to SIC principle. The above procedures are repeated until all OFDM-IM subblocks are demodulated.



\section{Simulation Results and Comparisons}

In this section, we present our theoretical as well as Monte Carlo simulation results for the MIMO-OFDM-IM scheme and make comparisons with the classical V-BLAST type MIMO-OFDM scheme for different type of detectors and system configurations.
The considered OFDM system parameters are summarized in Table III.

In Table IV, the decoding complexities of MIMO-OFDM and MIMO-OFDM-IM schemes are given in terms of total number of CMs performed per subcarrier for different type of detectors. As seen from Table IV, near-ML, MMSE-LLR and MMSE-LLR-OSIC detectors of MIMO-OFDM-IM have the same order decoding complexity compared to brute-force ML, MMSE and MMSE-OSIC detectors of classical MIMO-OFDM, respectively\footnote{In order to achieve the same spectral efficiency as that of MIMO-OFDM using the same $M$-ary constellation order, the spectral efficiency loss due to inactive subcarriers should be fully compensated by the bits carried with index modulation for MIMO-OFDM-IM. On the other hand, for the cases where MIMO-OFDM-IM is using higher order constellations, the increase in decoding complexity can be easily calculated from Table IV.}.

\begin{table}[!t]
	\begin{center}
		\caption{OFDM System Parameters}
		\begin{tabular}{|c||c|} \hline
			Number of Subcarriers $(N_F)$  & $512$ \\ 
			Subcarrier Spacing $(\Delta f)$ & $15$ kHz  \\ 
			Sampling Frequency $(f_s)$& $7.68$ MHz  \\ 
			Cyclic Prefix Length $(C_p)$ & $36$  \\ 
			MIMO Schemes $(T \times R)$& $2\times 2,2\times 4,4\times 4,4\times 8$  \\ 
			Modulation & BPSK, QPSK, $8/16/64$-QAM  \\
			Number of Multipaths $(L)$ & 10 \\ 
			\hline 
		\end{tabular} 
	\end{center}
	\vspace*{-0.3cm}
\end{table}

\begin{table*}[!t]
	\scriptsize
	\begin{center}
		
		\setlength{\extrarowheight}{1.5pt}
		\caption{Decoding Complexity Comparisons in Terms of Total Number of Complex Multiplications (CMs) Performed  per Subcarrier}
		\label{tab:Comp}
		\begin{tabular}[c]{|c|c|c|} \hline
			
			\textit{Detector} & \textit{MIMO-OFDM} & \textit{MIMO-OFDM-IM}  \\ \hline 
			Brute-Force ML & $R(T+1)M^T\sim \mathcal{O}(M^T)$ & $R(T+1)(CM^K)^T \sim \mathcal{O}(M^{KT})  $ \\ \hline
			Near-ML & n/a & $R(T+1)(M+1)^T \sim \mathcal{O}(M^T)$ \\ \hline
			MMSE (Simple)  & $T^3+2T^2 R + T(R+M) \sim \mathcal{O}(M)$ & $2T^3 + 5T^2 R + TR + CM^K \sim \mathcal{O}(M^K)  $\\ \hline
			MMSE-LLR  &  n/a & $2 T ^3 + 5T^2 R + T(R+M+1) \sim \mathcal{O}(M)$\\ \hline
			MMSE(-LLR)-OSIC  & $\!\!\!<\!T^4\!+ \! T^3(2R+3)\!+ \! 2T^2(R+1)\!+ \!T(2R+M) \sim \mathcal{O}(M)$ & $\!\!\!<\! T^4 \!+ \! T^3(2R+3) \!+ \! T^2 (4R+3) \!+ \! T(3R+M+1)\sim \mathcal{O}(M)$ \\ \hline
			Spectral Efficiency & $ \left( T N_F \log_2 M\right) /\left( N_F + C_p \right)   $ &  $ \left( NT \left( \lfloor \log_2\left(C\left(N,K\right)  \right)  \rfloor + K \log_2 M  \right)\right)  / \left( K \left(N_F + C_p \right)  \right)   $  \\ \hline
			
		\end{tabular}
	\end{center} 
	\vspace*{-0.3cm}
\end{table*}

\begin{figure}[!t]
	\begin{center}\resizebox*{11cm}{9cm}{\includegraphics{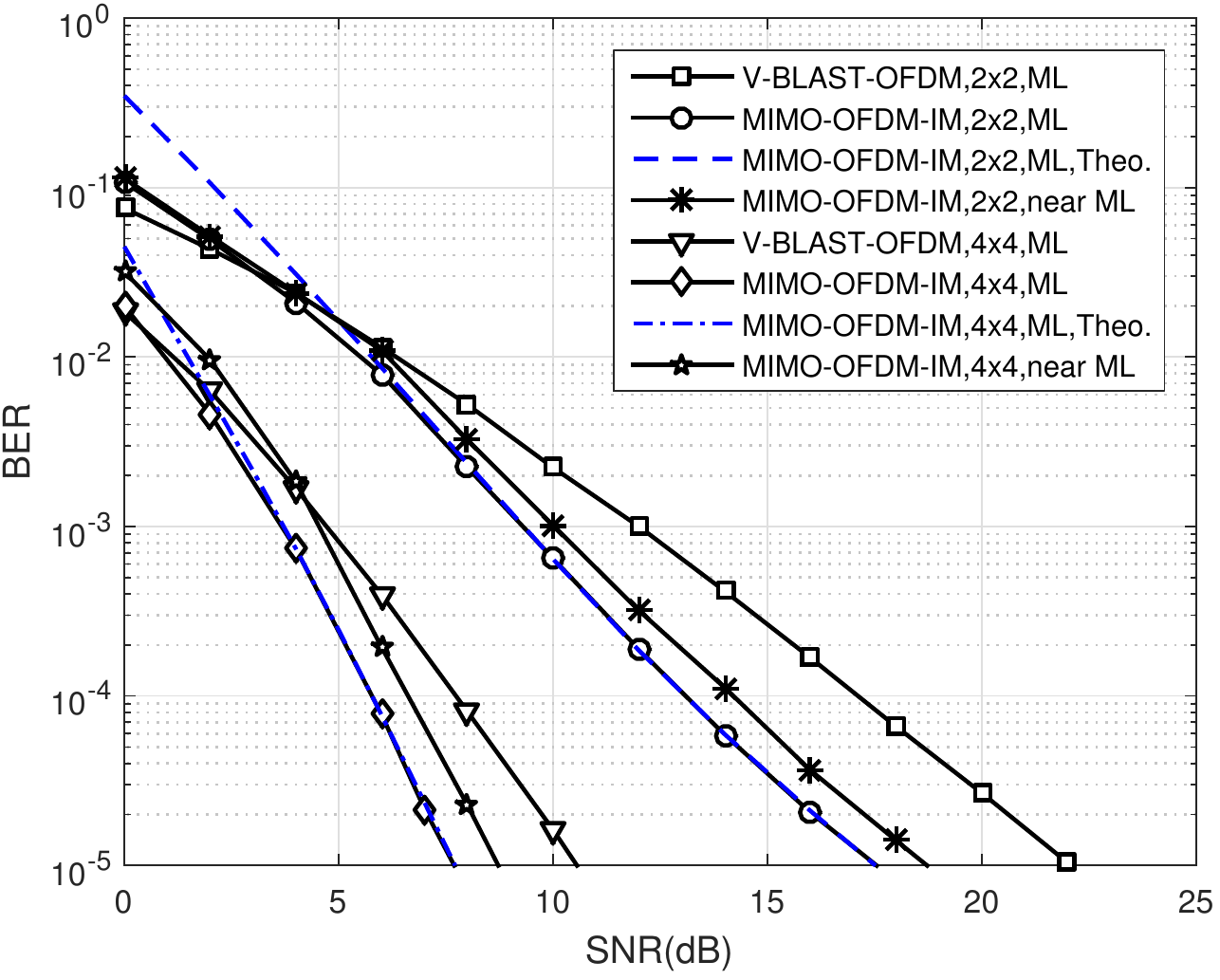}}
		\vspace*{-0.5cm}
		\caption{Performance comparison of V-BLAST-OFDM and MIMO-OFDM-IM $(N=4,K=2)$ for BPSK modulation $(M=2)$, ML/near-ML detection}
	\end{center}
	\label{fig:ML}
\end{figure}

In Fig. 3, we compare the BER performance of the MIMO-OFDM-IM scheme for $N=4,K=2$ (Table I) with classical V-BLAST-OFDM using ML detectors and BPSK modulation. Two MIMO configurations are considered: $2\times 2$ and $4\times 4 $, where both schemes achieve the same spectral efficiency values of $1.87$ and $3.74$ bits/s/Hz for these configurations, respectively. As seen from Fig. 3, using ML detection, a diversity order of $R$ is obtained for both schemes, and MIMO-OFDM-IM scheme provides considerable improvements in BER performance compared to classical V-BLAST-OFDM. It should be noted that the proposed near-ML detector outperformed by the brute-force ML detector by a small margin; however, it still exhibits considerably better BER performance than the classical V-BLAST-OFDM. We also observe that the derived theoretical upper in (\ref{eq:16}) becomes very tight with the computer simulation curves as the SNR increases.

\begin{figure}[!t]
	\begin{center}\resizebox*{13cm}{10cm}{\includegraphics{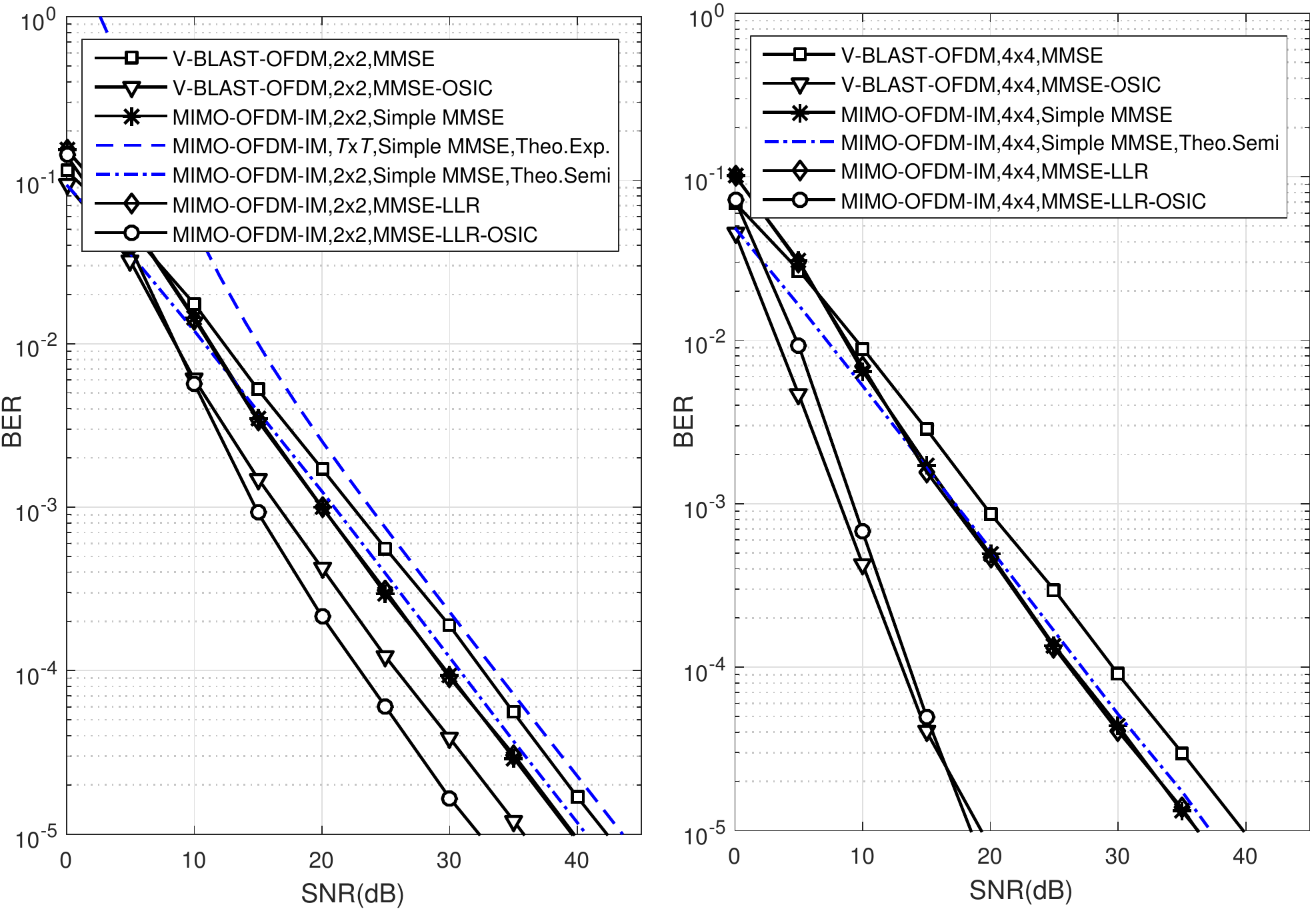}}
		\vspace*{-0.5cm}
		\caption{Performance comparison of V-BLAST-OFDM and MIMO-OFDM-IM ($N=4,K=3$ with reference look-up table) for QPSK modulation $(M=4)$, simple MMSE, MMSE-LLR, MMSE-LLR-OSIC detection}	
	\end{center}
	\vspace*{-0.4cm}
\end{figure}

In Fig. 4, we present the BER performance of the MIMO-OFDM-IM scheme for $N=4,K=3$ and classical V-BLAST-OFDM using MMSE type detectors and QPSK modulation. $2\times 2$ and $4\times 4 $ MIMO configurations are considered, where both schemes achieve the same spectral efficiency values of $3.74$ and $7.48$ bits/s/Hz for these configurations, respectively. As seen from Fig. 4, the simple MMSE and MMSE-LLR detectors provides almost the same BER performance for the MIMO-OFDM-IM scheme while they outperform classical V-BLAST-OFDM using MMSE detection. As expected, the theoretical ABEP curve which is based on the UPEP formula of (\ref{UPEP}) (exponentially distributed $Z_n$'s) provides a BER performance benchmark, while a much accurate ABEP curve can be obtained by using the semi-analytical UPEP calculation approach of (\ref{eq:UPEP_semi}). For comparison, the performance of OSIC based MMSE detectors are also shown in Fig. 4. As seen from Fig. 4, for the $2\times 2$ MIMO case, OSIC provides an SNR gain for both schemes; while for the $4\times 4$ MIMO case, the interference nulling becomes more dominant and a considerable improvement is observed in BER performance while MIMO-OFDM-IM still outperforms the reference V-BLAST-OFDM scheme with increasing SNR.

Fig. 5 presents the interesting trade-off provided by the MIMO-OFDM-IM scheme between BER performance and spectral efficiency  for a $4\times 4$ MIMO system with MMSE-LLR detection. For the selection of active indices, we use the combinatorial number theory method \cite{OFDM_IM}, where different $N$ and $K$ values are considered. As seen from Fig. 5, for the same spectral efficiency, the MIMO-OFDM-IM scheme with $M=8,N=16,K=13$ provides approximately $3$ dB better BER performance than the reference V-BLAST-OFDM scheme for a BER value of $10^{-5}$. On the other hand, as seen from Fig. 5, MIMO-OFDM-IM has the flexibility of adjusting the spectral efficiency by changing the number of active subcarriers $K$ in a subblock, and can achieve better or worse BER performance than the reference MIMO-OFDM-IM scheme with $11.2$ bits/s/Hz spectral efficiency. It should be noted that the BER performance of MIMO-OFDM-IM degrades when $64$-QAM is employed; however, a better BER performance than classical V-BLAST-OFDM is obtained for all other cases even with a higher spectral efficiency. The price paid for this improvement is the slightly increased signal processing at the receiver for the detection of data symbols and active indices.

Finally, we investigate the effects of a realistic channel model and imperfect channel estimation on the performance of the MIMO-OFDM-IM and make comparisons with classical V-BLAST-OFDM and Alamouti-OFDM in Fig. 6. Three MIMO configurations are considered: $2\times 4$ and $2/4\times 8 $, where all schemes achieve the same spectral efficiency values of $3.74$ and $7.48$ bits/s/Hz for these configurations, respectively. The considered CIR with $L=4$ taps is based on Extended Pedestrian A (EPA) model \cite{LTE_book} where the corresponding power delay profile is undersampled as in \cite{LTE_sampling} according with the sampling rate of the considered LTE-like system with $f_s=7.68$ MHz \cite{LTE_book}. In Table V, the normalized CIR values (with total power of unity) used in the simulations are given. In our computer simulations, each tap of the CIR is multiplied by an independent complex Gaussian r.v. with variance $0.5$ per dimension. We consider that the channel estimator at the receiver provides an estimate of the channel coefficients as \cite{TSC_estimation}
\begin{equation}
\mathbf{\hat{H}}_n^g=  \mathbf{H}_n^g + \mathbf{E}_n^g  
\end{equation}
where the elements of $ \mathbf{E}_n^g  $ follow $\mathcal{CN} \left(0, \sigma_e^2 \right) $ distribution. In this study, we assume that the power of channel estimation errors $\sigma_e^2$ decreases with the increasing SNR, i.e., $N_{0,F}$ and $ \sigma_e^2 $ are related via $Q= N_{0,F} / \sigma_e^2$. For the detection, the receiver considers the mismatched decision metrics where $\mathbf{\hat{H}}_n^g$ is used instead of $\mathbf{H}_n^g$ for all schemes.

As seen from Fig. 6, the BER performance of all systems considerably suffer from imperfect channel estimation $(Q=1)$ and similar levels of degradation are observed for all schemes. It is interesting to note that Alamouti-OFDM scheme achieves the best BER performance with increasing SNR values for the first configuration. On the other hand, its BER performance degrades considerably with increasing spectral efficiency due to the employment of a higher order constellation, and MIMO-OFDM-IM achieves the best BER performance for the second configuration. 

\begin{table}[!t]
	\begin{center}
		\caption{Channel Impulse Response Based on LTE-EPA Model}
		\begin{tabular}{|c|c|c|c|c|}
			\hline Linear Tap Gain & 0.7594  & 0.6486  & 0 & 0.0517 \\ 
			\hline Tap Delay $(ns)$ & $0$  & $130$  & $260$  & $390$  \\ 
			\hline 
		\end{tabular} 
	\end{center}
	\vspace*{-0.3cm}
\end{table}

\begin{figure}[!t]
	\begin{center}\resizebox*{11cm}{9cm}{\includegraphics{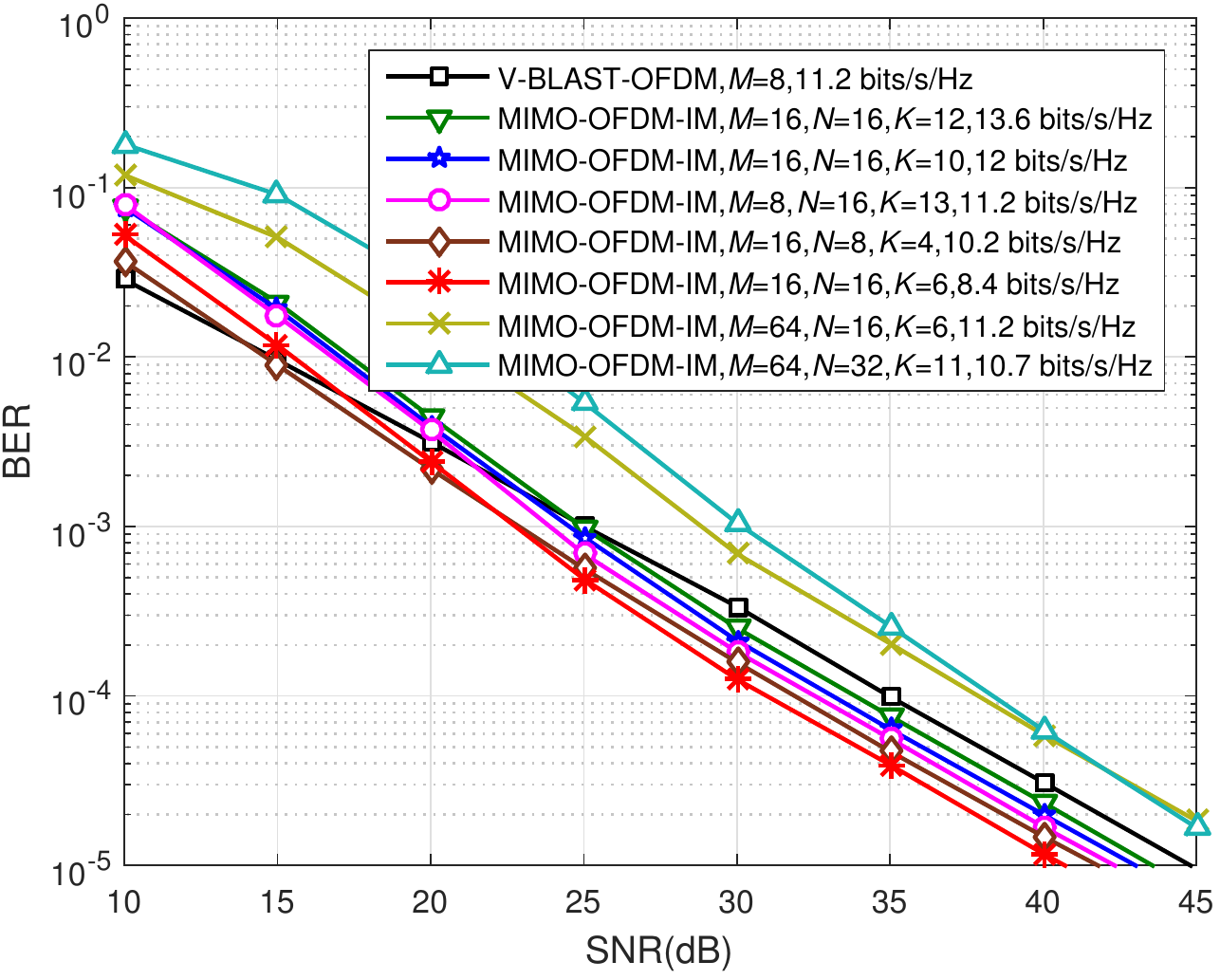}}
		\vspace*{-0.5cm}
		\caption{Performance comparison of V-BLAST-OFDM and MIMO-OFDM-IM for different $N$, $K$ and $M$ values and for a $4\times 4$ MIMO system with MMSE-LLR detection}
		\vspace*{-0.3cm}
	\end{center}
\end{figure}

\begin{figure}[!t]
	\begin{center}\resizebox*{13cm}{10cm}{\includegraphics{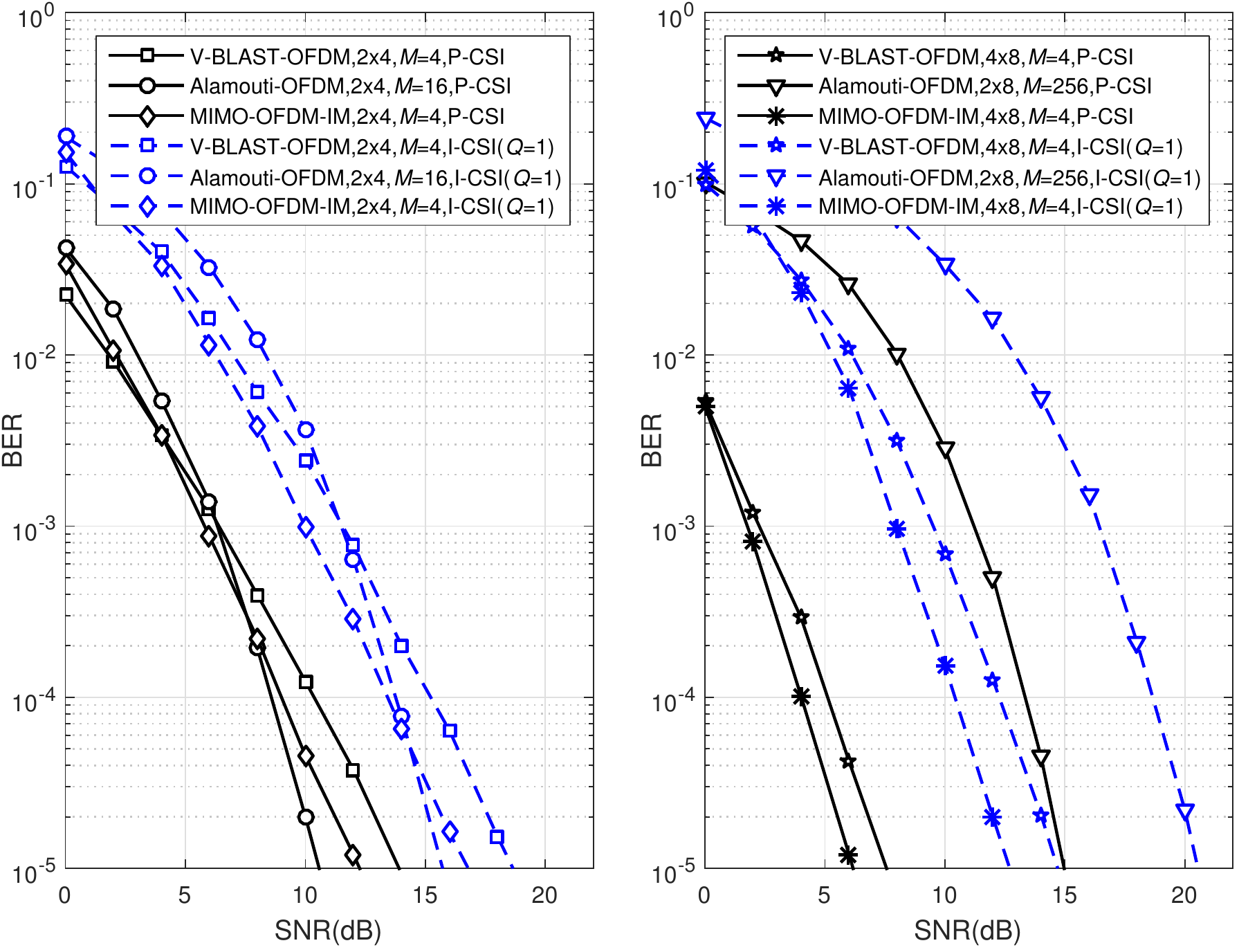}}
		\vspace*{-0.5cm}
		\caption{Performance comparison of V-BLAST-OFDM, Alamouti-OFDM and MIMO-OFDM-IM $(N=4,K=3)$ at $3.74$ and $7.48$ bits/sec/Hz for perfect and imperfect channel state information (P-CSI and I-CSI) cases with MMSE-LLR detection }
		\vspace*{-0.3cm}
	\end{center}
\end{figure}

\section{Conclusions}

In this study, the recently proposed MIMO-OFDM-IM scheme has been investigated for next generation 5G wireless networks. For the MIMO-OFDM-IM scheme, new detector types such as ML, near-ML, simple MMSE, MMSE-LLR-OSIC detectors have been proposed and their ABEP have been theoretically examined. It has been shown via extensive computer simulations that MIMO-OFDM-IM scheme provides an interesting trade-off between complexity, spectral efficiency and error performance compared to classical MIMO-OFDM scheme and it can be considered as a possible candidate for 5G wireless networks. The main features of MIMO-OFDM-IM can be summarized as follows: i) better BER performance, ii) flexible system design with variable number of active OFDM subcarriers and iii) better compatibility to higher MIMO setups. However, interesting topics such as diversity methods, generalized OFDM-IM cases, high mobility implementation and transmit antenna indices selection still remain to be investigated for the MIMO-OFDM-IM scheme.

\bibliographystyle{IEEEtran}
\bibliography{IEEEabrv,kitap_2016}

\vspace*{-7.2cm}

\begin{IEEEbiography}[{\includegraphics[width=1in,height=1.25in,clip,keepaspectratio]{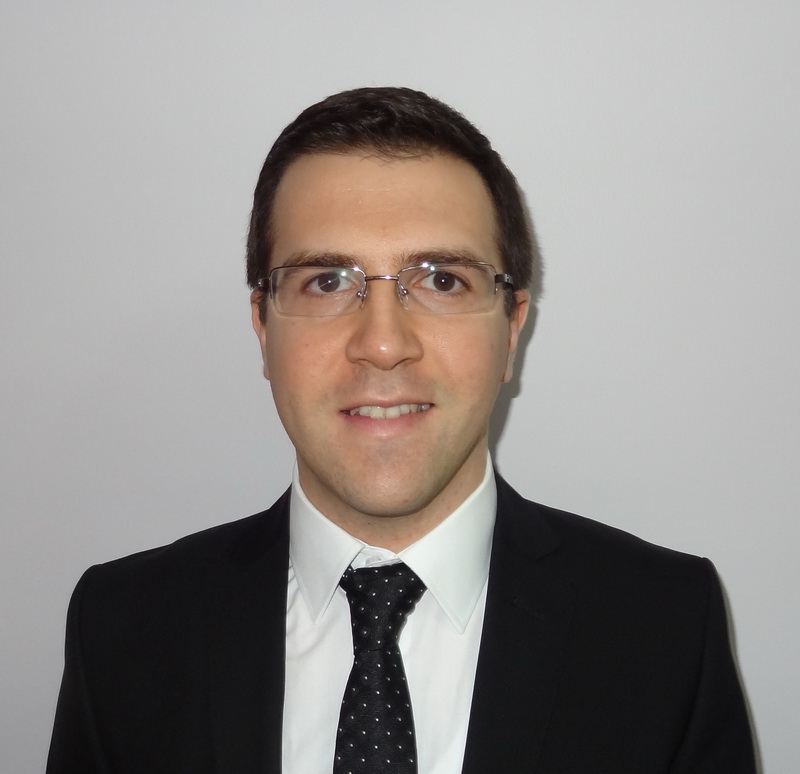}}]{Ertugrul Basar}
	(S'09, M'13, SM'16) was born in Istanbul, Turkey, in 1985. He received his B.S. degree with high honors from Istanbul University, Istanbul, Turkey, in 2007, and his M.S. and Ph.D. degrees from Istanbul Technical University, Istanbul, Turkey, in 2009 and 2013, respectively. He spent the academic year 2011-2012 at the Department of Electrical Engineering, Princeton University, New Jersey, USA. Currently, he is an assistant professor at Istanbul Technical University, Electronics and Communication Engineering Department and a member of Wireless Communication Research Group. He was the recipient of Istanbul Technical University Best Ph.D. Thesis Award in 2014 and won two Best Paper Awards. He is a regular Reviewer for various IEEE journals and served as a TPC Member for several conferences. His primary research interests include MIMO systems, index modulation, cooperative communications, OFDM, and visible light communications.
\end{IEEEbiography}


\end{document}